\documentclass[11pt]{amsart}
\usepackage[utf8]{inputenc}
\usepackage[margin=3cm]{geometry}
\usepackage{amssymb}
\usepackage{latexsym}
\usepackage{graphicx}
\usepackage[colorlinks]{hyperref}
\usepackage{comment}
\usepackage[english]{babel}
\usepackage{polski}
\usepackage{setspace}
\usepackage[all]{xy}
\usepackage{color}

\usepackage{subcaption}

\theoremstyle{plain}

\newtheorem{example}{Example}

\theoremstyle{definition}

\numberwithin{equation}{section}
\numberwithin{theorem}{section}

\numberwithin{equation}{section}
\numberwithin{theorem}{section}

\theoremstyle{definition}
\newcommand{\arrows}{\,\lower1pt\hbox{$\longrightarrow$}\hskip-.24in\raise2pt
             \hbox{$\longrightarrow$}\,}

\newcommand{\nsb}[1]{~\hspace{-4pt}_{^{[#1]}}}

\title{Geometry and solutions of an epidemic SIS model permitting fluctuations and quantization}

\begin{document}

\maketitle

\centerline{O\u{g}ul Esen$^{\dagger}$, Eduardo Fernández-Saiz$^{\ddagger}$, Cristina
Sard\'on$^{*}$,  Marcin Zaj\k{a}c$^{**}$}

\vskip 0.5cm

\centerline{Department of Mathematics$^{\dagger}$,}
\centerline{Gebze Technical University, 41400 Gebze, Kocaeli, Turkey.}
\centerline{oesen@gtu.edu.tr}
\vskip 0.2cm

\centerline{Departamento de Álgebra, Geometría y Topología$^{\ddagger}$}
\centerline{Universidad Complutense de Madrid, Pza. Ciencias 3, E-28040 Madrid, Spain}
\centerline{eduardfe@ucm.es}
\vskip 0.2cm

\centerline{Department of Applied Mathematics$^*$} 
\centerline{Universidad Polit\'ecnica de Madrid}
\centerline{C/ Jos\'e Guti\'errez Abascal, 2, 28006, Madrid. Spain}
\centerline{mariacristina.sardon@upm.es}
\vskip 0.2cm

\centerline{Department of Mathematical Methods in Physics$^{**}$}
\centerline{University of Warsaw, ul. Pasteura 5, 02-093 Warsaw, Poland.}
\centerline{marcin.zajac@fuw.edu.pl}

\begin{abstract}

Some recent works reveal that there are models of differential equations for the mean and variance of infected individuals that reproduce the SIS epidemic model at some point. 
This stochastic SIS epidemic model can be interpreted as a Hamiltonian system, therefore we wondered if it could be geometrically handled through the theory of Lie--Hamilton systems, and this happened to be the case. The primordial result is that we are able to obtain a general solution for the  stochastic/ SIS-epidemic model (with fluctuations) in form of a nonlinear superposition rule that includes particular stochastic solutions and certain constants to be related to initial conditions of the contagion process. The choice of these initial conditions will be crucial to display the expected behavior of the curve of infections during the epidemic. We shall limit these constants to nonsingular regimes and display graphics of the behavior of the solutions. As one could expect, the increase of infected individuals follows a sigmoid-like curve.

Lie--Hamiltonian systems admit a quantum deformation, so does the stochastic SIS-epidemic model. We present this generalization as well. If one wants to study the evolution of an SIS epidemic under the influence of a constant heat source (like centrally heated buildings), one can make use of quantum stochastic differential equations coming from the so-called quantum deformation.

\end{abstract}

\tableofcontents

\onehalfspacing

\setlength{\parindent}{0cm}
\setlength{\parskip}{0.5cm}
\section{Introduction}

Epidemic models try to predict the spread of an infectious disease afflicting a specific population, see for example \cite{Brauer,Mi83,Mu07}. These models are rooted in the works of Bernoulli in the 18th century, when he proposed a mathematical model to defend the practice of inoculating against smallpox \cite{smallpox}. This was the start of germ theory.

At the beginning of the 20th century, the emergence of compartmental models was starting to develop. Compartmental models are deterministic models in which the population is divided into compartments, each representing a specific stage of the epidemic. For example, $S$ represents the susceptible individuals to the disease, $I$ designates the infected individuals, whilst $R$ stands for the recovered ones. The evolution of these variables in time is represented by a system of ordinary differential equations whose independent variable, the time, is denoted by $t$. 
Some of these first models are the Kermack--McKendrick \cite{KK} and the Reed--Frost \cite{Ab52} epidemic models, both describing the dynamics of healthy and infected individuals among other possibilities. There are several types of compartmental models \cite{Harko, Hethcote, Miller}, as it can be the SIS model, in which after the infection the individuals do not adquire immunity, the SIR model, in which after the infection the individuals adquire immunity, 
the SIRS model, for which immunity only lasts for a short period of time, the MSIR model, in which infants are born with immunity, etc. In this present work our focus is on the SIS model. 

\textbf{The SIS model.} The susceptible-infectious-susceptible (SIS) epidemic model assumes a population of size $N$ and one single disease disseminating.
The infectious period extends throughout the whole course of the disease until the recovery of the patient with two possible states, either infected or susceptible. This implies that there is no immunization in this model. 
In this approach the only relevant variable is the instantaneous density of infected individuals $\rho=\rho(\tau)$ depending on the time parameter $\tau$, and taking values in $[0,1]$. The density of infected individuals decreases with rate $\gamma \rho$, where $\gamma$ is the recovery rate, and the rate of growth of new infections is proportional to $\alpha \rho(1-\rho)$, where the intensity of contagion is given by the transmission rate $\alpha$. These two processes are modelled through the compartmental equation
\begin{equation}\label{compsis1}
\frac{d \rho}{d\tau}=\alpha \rho(1-\rho)-\gamma \rho.
\end{equation}
One can redefine the timescale as $t := \alpha \tau$ and introduce the constant $\rho_0:=1-\gamma/\alpha$, so we can rewrite \eqref{compsis1} as
\begin{equation}\label{sismodel0}
\frac{d\rho}{dt}= \rho(\rho_0-\rho).
\end{equation}
Clearly, the equilibrium density is reached if $\rho=0$ or $\rho=\rho_0$.
Although compartmental equations have proven their efficiency for centuries, they are still based on strong hypotheses. For example, the SIS model works more efficiently under the random mixing and large population assumptions. The first assumption is asking homogeneous mixing of the population, that is, individuals contact with each other randomly and do not gather in smaller groups, as abstaining themselves from certain communities. This assumption is nevertheless rarely justified. The second assumption is the rectangular and stationary age distribution, which means that everyone in the population lives to an age $L$, and for each age up to $L$, which is the oldest age, there is the same number of people in each subpopulation. This assumption seems feasible in developed contries where there exists very low infant mortality, for example, and a long live expectancy. Nonetheless, 
it looks reasonable to implement probability at some point to permit random variation in one or more inputs over time. Some recent experiments provide evidence that temporal fluctuations can drastically alter
the prevalence of pathogens and spatial heterogeneity also introduces an extra layer of complexity as it can delay the pathogen transmission \cite{duncan, real}.

\textbf{The SIS model with fluctuations.} 
It is needless to point out that fluctuations should be considered in order to capture the spread of infectious diseases more closely. Nonetheless, the introduction of these fluctuations is not trivial. One way to account for fluctuations is to consider stochastic variables. On the other hand, it seems that in the case of SIS models there exist improved differential equations for the mean and variance of infected individuals. 
Recently, in \cite{NakamuraMartinez}, the model assumes the spreading of the disease as a Markov chain in discrete time in which at most one single recovery or transmission occurs in the duration of this infinitesimal interval.
As a result \cite{kiss}, the first two equations for instantaneous mean density of infected people $\langle\rho\rangle$ and the variance $\sigma^2=\langle\rho^2\rangle-\langle\rho\rangle^2$ are 
\begin{equation}\label{sismodel1}
\begin{split}
\frac{d\langle\rho\rangle}{dt}&=\langle\rho\rangle \left( \rho_0-\langle\rho\rangle \right)-\sigma^2 ,\\
\frac{d\sigma^2}{dt}&=2\sigma^2 \left(\rho_0-\langle\rho\rangle\right)-\Delta_3 -\frac{1}{N}\langle\rho (1-\rho)\rangle+\frac{\gamma}{N\alpha}\langle\rho\rangle
\end{split}
\end{equation}
where $\Delta_3 =\langle\rho^3\rangle-\langle\rho\rangle^3$. This system  finds excellent agreement with empirical data \cite{NakamuraMartinez}. 
Equations \eqref{sismodel0} and \eqref{sismodel1} are equivalent when $\sigma$ becomes irrelevant compared to $\langle\rho\rangle$. Therefore, a generalization of compartmental equations only requires mean and variance, neglecting higher statistical moments. The skewness coefficient vanishes as a direct consequence of this assumption, so that $\Delta_3 := 3\sigma^2\langle\rho \rangle$. For a big number of individuals $(N\gg 1)$, the resulting equations are
\begin{equation}\label{sislabellog}
\begin{split}
\frac{d \ln{\langle\rho\rangle}}{d t}&=\rho_0-\langle\rho\rangle-\frac{\sigma^2}{\langle\rho\rangle},\\
\frac{1}{2}\frac{d \ln{\sigma^2}}{d t}&=\rho_0-2\langle\rho\rangle.
\end{split}
\end{equation}
The system right above can be obtained from a stochastic expansion as it is given in \cite{vilar}, as well.  

\textbf{Hamiltonian character of the model \eqref{sislabellog}.} The investigation of the geometric and/or the algebraic foundations of a system permits to employ several powerful techniques of geometry and algebra while performing the qualitative analysis of the system \cite{FM,Ar13,LiMa}. This even results in an analytical/general solution of the system in our case. For example, we cite \cite{LeAn03} and \cite{NuLe04} for Lie symmetry approach to solve the classical SIS model. Particularly, the Hamiltonian analysis of a system plays an important role in the geometrical analysis of a given system. For instance, in compartmental theories we can cite an early work \cite{Nu90} that classifies the SIR model as bihamiltonian. We also refer to \cite{EsGu18} for conformal Hamiltonian analysis of the Kermack-McKendrick Model.  In a very recent study \cite{Ballesteros2020}, it is shown that all classical epidemic compartmental models admit Hamiltonian realizations. 

In \cite{NakamuraMartinez}, the SIS system \eqref{sislabellog} involving fluctuations has been recasted in Hamiltonian form in the following way: the dependent variables are the mean $\langle\rho\rangle$ and the variance $\sigma^2$, and they both depend on time. Then, we define the dynamical variables $q=\langle\rho\rangle$ and  $p=1/\sigma$, so the system \eqref{sislabellog} turns out to be 
\begin{equation}\label{sismodel3}
\begin{split}
\frac{dq}{dt}&=q\rho_0-q^2-\frac{1}{p^2},
\\
\frac{dp}{dt}&=-p\rho_0+2pq.
\end{split}
\end{equation}
We employ the abbreviation SISf for system \eqref{sismodel3} to differentiate it from the classical SIS model in \eqref{sismodel0}. The letter ``f" accounts for ``fluctuations".

We have computed the general solution to this system, finding a more general solution than the one provided by Nakamura and Mart\'inez in \cite{NakamuraMartinez}. Indeed, we have found obstructions in their model solution. We shall comment this in the last section gathering all our new results.

Our general solution for this system reads:
\begin{equation} \label{solsismodel3}
\begin{split}
q(t)&=\frac{\rho e^{\rho t}(C_1\rho^2-4)e^{\rho t}+2C_1C_2\rho^2}{(C_1^2\rho^2-4)e^{2\rho t}+4C_2\rho^2(C_1e^{\rho t}+C_2)},\\
p(t)&=C_1+\frac{C_1^2\rho^2-4}{4\rho^2-C_2}+C_2e^{-\rho t}.
\end{split}
\end{equation}
In order to develop a geometric theory for this system of differential equations, we need to choose certain particular solutions that we shall make use of. Here we present three different choices and their corresponding graphs according to the change of variables $q=\langle \rho\rangle $ and $p=1/\sigma$.

\begin{figure}[htb]
\centering
\begin{subfigure}[b]{0.45\linewidth}
\includegraphics[width=\linewidth]{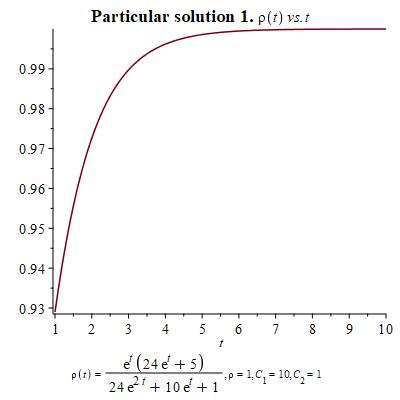}
\end{subfigure}
\begin{subfigure}[b]{0.45\linewidth}
\includegraphics[width=\linewidth]{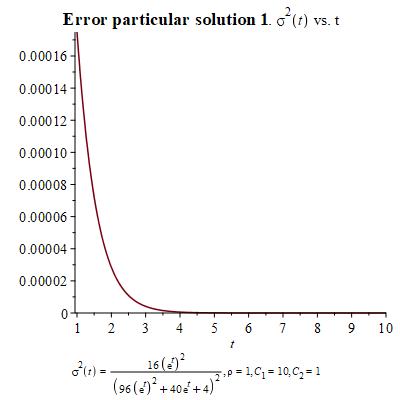}
\end{subfigure}
\caption{The first particular solution}
\label{PS1}
\end{figure}

\newpage

\begin{figure}[htb]
\centering
\begin{subfigure}[b]{0.45\linewidth}
\includegraphics[width=\linewidth]{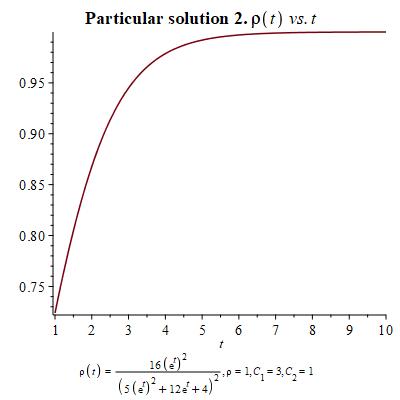}
\end{subfigure}
\begin{subfigure}[b]{0.45\linewidth}
\includegraphics[width=\linewidth]{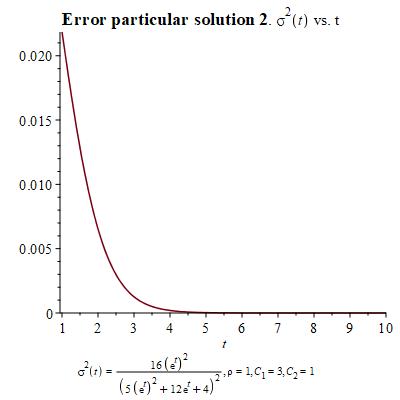}
\end{subfigure}
\caption{The second particular solution}
\label{PS2}
\end{figure}

\begin{figure}[htb]
\centering
\begin{subfigure}[b]{0.45\linewidth}
\includegraphics[width=\linewidth]{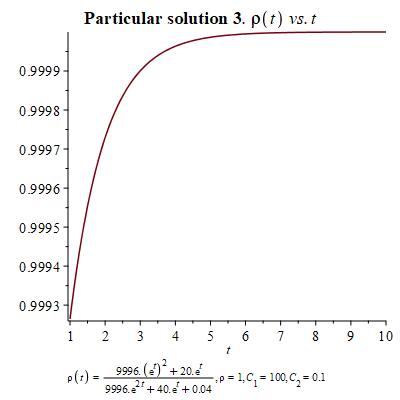}
\end{subfigure}
\begin{subfigure}[b]{0.45\linewidth}
\includegraphics[width=\linewidth]{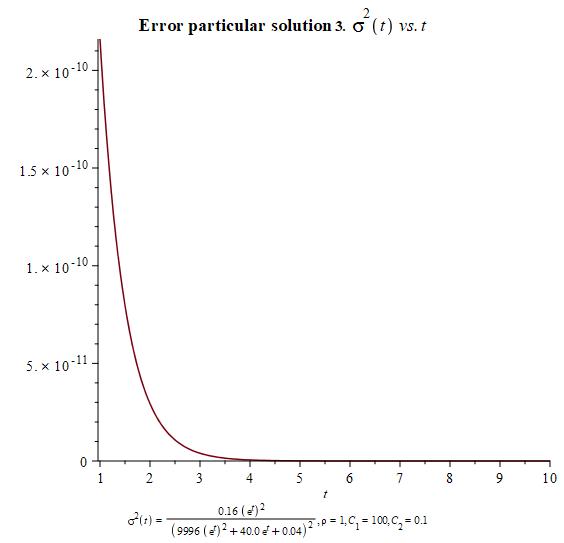}
\end{subfigure}
\caption{The third particular solution}
\label{PS3}
\end{figure}

Let us turn now to interpret these equations geometrically on a symplectic manifold.
The symplectic two-form $\omega=dq\wedge dq$ is a canonical skew-symmetric tensorial object in two-dimensions. For a chosen (real-valued) Hamiltonian function $h=h(q,p)$, the dynamics is governed by a Hamiltonian vector field $X_h$ defined through the Hamilton equation
\begin{equation} \label{Ham-Eq}
\iota_{X_h}\omega=dh,
\end{equation}
where $\iota_{X_h}$ is the contraction operator in the tensor algebra, and $dH$ is the (exterior) derivative of $h$. In terms of the coordinates $(q,p)$, the Hamilton equations \eqref{Ham-Eq} become
\begin{equation} \label{loc-Ham-Eq}
\frac{d q} {dt}=\frac{\partial h}{\partial p},\qquad \frac{d p} {dt}=-\frac{\partial h}{\partial q}.
\end{equation}
It is possible to realize that the SISf system (\ref{sismodel3}) is a Hamiltonian system since it fulfills the Hamilton equations \eqref{Ham-Eq}. To see this, consider the Hamiltonian function
\begin{equation}\label{hamiltoniansis}
h=qp\left(\rho_0-q\right)+\frac{1}{p}.
\end{equation}
and substitute it into \eqref{loc-Ham-Eq}. A direct calculation will lead us to (\ref{sismodel3}). The skew-symmetry of the symplectic two-form implies that the Hamiltonian function is constant all along the motion. In classical mechanics, where the Hamiltonian is taken to be the total energy, this corresponds to the conservation of energy. 

\textbf{The goal of the present work.} Our focus is the SISf model permitting fluctuations given in \eqref{sismodel3}. We shall perform both the Lie analysis and the Lie-Hamilton analysis of this system in order to explore the geometric/algebraic foundations of the model as well as to arrive at the analytical general solution. Further, we shall introduce a deformation (the so-called quantum deformation) of the model \eqref{sismodel3} to make it applicable to more complicated issues. 

To get involved in the mathematical theories needs some special emphasis or/and some expertise. For that, we shall try to present both the theoretical and the practical results in the most accessible forms for a general audience  as far as we can.  Our plan to achieve this is to exhibit itinerary maps, graphs, and a comprehensive appendix.  

The paper is organized into three main sections, and an appendix. 
Sections \ref{Lie-Sec} and \ref{Lie-Ham-Sec} concern the Lie and the Lie-Hamilton analysis of the SISf model \eqref{sismodel3}, respectively. Section \ref{Sec-quant} proposes the so-called quantum deformation of the model and Appendix \ref{fund} is a brief review of some algebraic constructions, such as Lie algebras, Poisson algebras, Lie coalgebras and Poisson Hopf algebras, that we refer to in the main body of the paper.




\newpage

\section{Lie Analysis of the SISf model} \label{Lie-Sec}

A (nonautonomous) system of ordinary differential equations (ODEs) can be written as a time-dependent vector field and vice versa. Such system is called Lie system if the associated time-dependent vector field takes values in a finite-dimensional Lie algebra of vector fields \cite{CGM,CGM07}. Equivalently, one can also define a Lie system as a system of ODEs admitting a(nonlinear) superposition principle, that is, a map allowing us to express the general solution of the system of ODEs in terms of a family of particular solutions and a set of constants related to initial conditions. Accordingly, in this section, we shall show that the SISf model \eqref{sismodel3} is a Lie system. For the completeness of this work, we start with a brief summary of some basic notions concerning the theory of Lie systems. We refer to Appendix \ref{App-LA} for the definitions of Lie (sub)algebras, and to Appendix \ref{App-TB} for the definitions of (co)tangent bundle and vector fields. For some further reading on Lie systems especially about its relevant role in
Physics, Mathematics, Biology, Economics, and other fields of research, see \cite{CL,LS,sardon}  and extensive reference lists there in. 

\subsection{Time-dependent vector fields}~

Let $N$ be a manifold. A time-dependent vector field on $N$ is a differentiable mapping
\begin{equation}
X:\mathbb{R}\times N\longrightarrow TN, \qquad  \tau_N\circ X=pr_2
\end{equation}
where $pr_2$ is the projection to the second factor in $\mathbb{R}\times N$. In this regard, we can consider a time-dependent vector as a set (family) of standard vector fields $\{X_t\}_{t\in\mathbb{R}}$, depending on a single parameter. The converse of this assertion is also true, that is a set of vector fields depending smoothly on a single parameter can be written as a time-dependent vector field \cite{CL}. We plot the following diagram to summarize the discussions.
\begin{equation}
\xymatrix{
    & TN \ar[d]^{\tau_N} \\
    \mathbb{R}\times N \ar[ru]^{X} \ar[r]_{pr_2}       & N }
\end{equation}
A time dependent vector field determines a non-autonomous ODE system 
\begin{equation}\label{genLiesys1}
 \frac{dx}{dt}=X(t,x),\qquad x\in N.
\end{equation}
  The suspension of a time-dependent vector field $X$ on $N$ is the vector field on $\mathbb{R}\times N$ defined to be $ X+\partial/\partial t$, where $t$ is the global coordinate for $\mathbb{R}$. An integral curve of a time-dependent vector field $X$ is an integral curve of the suspension \cite{FM}. Conversely,
every system of first-order differential equations in normal form describes the
integral curves of a  unique time-dependent vector field.

Consider the case that $N$ is a two dimensional manifold with local coordinates $(q,p)$, and the extended space $\mathbb{R}\times N$ with induced coordinates $(t,q,p)$. A time dependent vector field is written as 
\begin{equation} \label{time-dept-vf-loc}
X=k(t,q,p)\frac{\partial}{\partial q}+g(t,q,p)\frac{\partial}{\partial q}
\end{equation}
where $k$ and $g$ are real valued functions on $\mathbb{R}\times N$. In this coordinate system, the dynamical equations governed by $X$ determine a system of nonautonomous ordinary differential equations
\begin{equation}
\dot{q}=k(t,q,p), \qquad \dot{p}=g(t,q,p).
\end{equation}
Notice that, the SISf model \eqref{sismodel3} fits this picture with even the generalization $\rho_0=\rho(t)$. We shall turn to this discussion in the following subsections.

\textbf{Vessiot-Guldberg algebra.} Start with a time-dependent vector field $X$, and determine the set $\{X_t\}_{t\in\mathbb{R}}$ of vector fields. We denote the smallest Lie subalgebra containing the set $\{X_t\}_{t\in\mathbb{R}}$ by $V^X$. 
The associated distribution $\mathcal{D}^X$ for a time-dependent vector field $X$ is the generalized  distribution spanned by the vector
fields in $V^X$ that is
\begin{equation}
\mathcal{D}^X_x=\{Y_x\mid Y\in V^X\}\subset T_xN.
\end{equation}
The associated codistribution $(\mathcal{D}^X)^\circ$ for a time-dependent vector field $X$ is then defined to be the annihilator of $\mathcal{D}^X$ that is
\begin{equation}
(\mathcal{D}^X_x)^\circ=\{\vartheta\in T_x^*N\mid \vartheta(Z_x)=0,\quad \forall
 Z_x\in \mathcal{D}_x^X\}.
\end{equation} 
is a subbundle of the cotangent bundle $T^*N$. Our interest relies in a distribution $\mathcal{D}^X$ determined by a
finite-dimensional $V^X$ and hence $\mathcal{D}^X$ 
becomes integrable on the whole $N$. In this case, $V^X$ is called Vessiot-Guldberg Lie algebra. 
It is worth noting that even in this case, $(\mathcal{D}^X)^\circ$ does not need to be a {
differentiable codistribution.

A function $f$ on an open neighbourhood $U$ 
is a local $t$-independent constant of  
motion for a system $X$ if and only if $df(x)$ in $(\mathcal{D}^X_x)^\circ|_U$ for all $x$ in $N$.
This statement reads that (locally defined) time-independent
constants of motion of time-dependent vector fields are determined
 by (locally defined) exact one-forms taking values in the associated
codistribution. Therefore, $(\mathcal{D}^X)^\circ$ is a crucial object in the calculation of such constants of motion for a system $X$. 
 
\subsection{Lie Systems} ~
 
Let $X$ be a time-dependent vector field 
defined on a manifold $N$. A superposition rule depending on $m$ particular solutions of $X$  is a function 
\begin{equation}
\phi:N^m\times N\rightarrow N, \qquad x=\phi(x_{(1)},\dots, x_{(m)},\lambda)
\end{equation}  such that the general
solution $x(t)$ of $X$ can be written as $x(t)=\phi(x_{(1)}(t),\dots, x_{(m)}(t),\lambda)$
for any generic family $x_{(1)}(t), \dots , x_{(m)}(t)$ of particular solutions. Here, $\lambda$ is a point of $N$ related with the initial conditions. A system of equations admitting superposition rule is called Lie system.

\textbf{Lie--Scheffers Theorem.}
A modern statement
of this result is described in \cite{CGM07}.
A first-order system \eqref{genLiesys1}  
admits a superposition rule if and only if the associated time-dependent vector field $X$ can be written as
 \begin{equation}\label{genLiesys}
X_t={{\sum_{\alpha=1}^r}}b_\alpha(t)X_\alpha
\end{equation}
for a certain family $b_1(t),\dots,b_r(t)$  of time-dependent functions and a
family  $X_1,\dots,X_r$ of vector fields  on N spanning 
an $r$-dimensional real Lie subalgebra of vector fields. We refer to \cite{LS} for details, see  \cite{zelikin,reid} for some further discussions and the first examples. The Lie--Scheffers 
Theorem yields that a
system $X$ admits a superposition rule if and only if $V^X$ is
finite-dimensional, \cite{CL}.

\textbf{An itinerary map to derive superposition rules.}
General solutions of Lie systems can also be investigated through superposition rules. There exist various procedures to derive them \cite{w1,winter1}, but we hereafter use the method
devised in \cite{CGM07}, which is based on the notion of {diagonal prolongation} \cite{CL}. Let us denote the $m+1$-times Cartesian product of $N$ by itself as $N^{(m+1)}$. We denote by
\begin{equation}
{\rm pr}:N^{(m+1)}\longrightarrow N, \qquad (x_{(0)},x_{(1)},\ldots,x_{(m)})\mapsto x_{(0)}
\end{equation}
the projection onto the first factor. Given a time-dependent vector field $X$ on $N$, the diagonal prolongation $\widetilde X$ of $X$ to the product space $N^{(m+1)}$ is a unique time-dependent vector field on $N^{(m+1)}$ projecting to $X$ by the projection ${\rm pr}$ that is ${\rm pr}_*\widetilde X_t=X_t$ for all $t$. We also require $\widetilde X$ to be invariant under the permutations $x_{(i)}\leftrightarrow x_{(j)}$ with $i,j=0,\dots,m$.
The procedure to determine superposition rules described in \cite{CGM07} goes as follows:

\textbf{Step 1.} Take a basis $X_1,\dots,X_r$ 
of a Vessiot--Guldberg Lie algebra associated with the Lie system $X$.

\textbf{Step 2.}  Choose the minimum integer $m$ so that 
the diagonal prolongations $\widetilde X_1,\ldots,\widetilde X_r,$ of $X_1,\dots,X_r$ to $N^{m}$ are linearly independent at a generic point. 

\textbf{Step 3.}  Obtain, for instance, by the method of characteristics, $n$ functionally independent first-integrals $F_1,\ldots,F_n$ common to all the diagonal prolongations, $\widetilde X_1,\ldots,\widetilde X_r,$ to $N^{(m+1)}$. We require such functions to satisfy
\begin{equation*}
\frac{\partial (F_1,\dots,F_n)}{\partial \left((x_1)_{(0)},\dots,(x_n)_{(0)}\right)}\neq 0.
\end{equation*}
Assume that these integrals take certain constant values, i.e., $F_i=k_i$ where $i=1,\ldots,n$, and
employ these equalities to express the variables $(x_1)_{(0)},\dots,(x_n)_{(0)}$ in terms of the variables of the other copies of $N$ within $N^{(m+1)}$ and the constants $k_1,\ldots,k_n$. 

The obtained 
 expressions constitute a superposition rule in terms of any generic family of $m$ particular solutions and $n$ constants. Let us apply these steps in the realm of the SISf model. 

\subsection{The SISf is a Lie system}~

The model \eqref{sismodel3} can be generalized to a model represented by a time-dependent vector field 
\begin{equation}
X_t=\rho_0(t)X_1+X_2
\end{equation}
where the constitutive vector fields are computed to be
\begin{equation} \label{sisliesystem-}
X_1=q\frac{\partial}{\partial q}-p\frac{\partial}{\partial p},\quad X_2=\left(-q^2-\frac{1}{p^2}\right)\frac{\partial}{\partial q}+2qp\frac{\partial}{\partial p}.
\end{equation}
The generalization comes from the fact that $\rho_0(t)$ is no longer a constant, but it can evolve in time. Let us apply the steps introduced in the previous section one by one to arrive at the general solution. 

\textbf{Step 1.} For the vector fields in \eqref{sisliesystem-}, a direct calculation shows that the Lie bracket 
\begin{equation}\label{sisliesystem}
[X_1,X_2]=X_2
\end{equation}
is closed within the Lie algebra. This implies that the SISf model \eqref{sismodel3} is a Lie system. The Vessiot-Guldberg algebra spanned by $X_1, X_2$ is an imprimitive Lie algebra of type $I_{14}$ according to the classification presented in \cite{Ballesteros2}. 

\textbf{Step 2.} If we copy the configuration space twice, we will have four degrees of freedom $(q_1,p_1,q_2,p_2)$ and we will archieve precisely two first-integrals in vinicity of the Fr\"obenius theorem.
A first-integral for $X_t$ has to be a first-integral for $X_1$ and $X_2$ simultaneously. We define the diagonal prolongation $\widetilde{X}_1$ of the  vector field $X_1$ in the decomposition \eqref{sisliesystem}. Then we look for a first integral $F_1$ such that $\widetilde{X}_1[F_1]$ vanishes identically. Notice that if $F_1$ is a first-integral of the vector field $\widetilde{X}_1$ then it is a first integral of $\widetilde{X}_2$ due to the commutation relation.
For this reason, we start by integrating the prolonged vector field 
\begin{equation}
\widetilde{X}_1=q_1\frac{\partial}{\partial q_1}+q_2\frac{\partial}{\partial q_2}-p_1\frac{\partial}{\partial p_1}-p_2\frac{\partial}{\partial p_2}
\end{equation}
through the following characteristic system
\begin{equation}
\frac{dq_1}{q_1}=\frac{dq_2}{q_2}=\frac{dp_1}{-p_1}=\frac{dp_2}{-p_2}.
\end{equation}
Fix the dependent variable $q_1$ and obtain a new set of dependent variables, say $(K_1,K_2,K_3)$, which are computed to be 
\begin{equation}\label{kchange}
K_1=\frac{q_1}{q_2},\qquad K_2=q_1p_1,\qquad K_3=q_1p_2.
\end{equation}

\textbf{Step 3.} This induces the following basis in the tangent space
\begin{equation}
\begin{split}
\frac{\partial}{\partial K_1}=q_2 \frac{\partial }{\partial q_1}-\frac{q_2 p_1}{q_1} \frac{\partial }{\partial p_1}- \frac{q_2 p_2}{q_1} \frac{\partial }{\partial p_1}, \qquad \frac{\partial}{\partial K_2}=\frac{1}{q_1}\frac{\partial}{\partial p_1}, \qquad
\frac{\partial}{\partial K_3}=
\frac{1}{q_1}\frac{\partial}{\partial p_2}.
\end{split}
\end{equation}
provided that $q_1$ is not zero. Introducing the coordinate changes exhibited in \eqref{kchange} into the diagonal projection  $\widetilde{X}_2$ of the vector field $X_2$, we arrive at the following expression
\begin{equation}
\begin{split}
\widetilde{X}_2=&\left(2K_1-\left(1+\frac{1}{K_1^2}\right)\right)\frac{\partial}{\partial K_1}+\left(\left(\frac{1}{K_2^2}+\frac{1}{K_3^2}\right)K_2^2-\left(1+\frac{1}{K_1^2}\right)K_2\right)\frac{\partial}{\partial K_2}\nonumber\\
&+\left(2\frac{K_3}{K_2}-\left(1+\frac{1}{K_1^2}\right)K_3\right)\frac{\partial}{\partial K_3}.
\end{split}
\end{equation}
To integrate the system once more, we use the method of characteristics again and obtain
\begin{equation}\label{logsys}
\frac{d\ln{|K_1|}}{1-\frac{1}{K_1^2}}=\frac{d\ln{|K_2|}}{\frac{1}{K_2}+\frac{K_2}{K_3}-1-\frac{1}{K_1^2}}=\frac{d\ln{|K_3|}}{\frac{2}{K_2}-\left(1+\frac{1}{K_1^2}\right)}.
\end{equation}

\textbf{Exact solution.}
We obtain two first integrals by integrating in pairs $(K_1,K_2)$ and $(K_1,K_3)$, where we have fixed $K_1$. After some cumbersome calculations we obtain
\begin{equation} \label{KKK}
K_2=\frac{K_1\left(4k_2^2 K_1^2+4k_1k_2K_1+k_1^2-4\right)}{2(K_1+1)(K_1-1)k_2(2k_2K_1+k_1)}, \qquad 
K_3=\frac{K_1\left(k_2K_1^2+k_1K_1+\frac{k_1^2-4}{4k_2}\right)}{(K_1+1)(K_1-1)}.
\end{equation}
By substituting back the coordinate transformation  \eqref{kchange} into the solution \eqref{KKK} (please notice the difference between capitalized constants $(K_1,K_2,K_3)$ and lower case constants $(k_1,k_2)$, we arrive at the following implicit equations
\begin{equation}\label{exactsol}
\begin{split}
q_1&=-\frac{q_2\left(k_1k_2\pm \sqrt{4k_2^2p_2^2q_2^2+k_1^2k_2p_2q_2-4k_2^3p_2q_2-4k_2p_2q_2+4k_2^2}\right)}{2k_2(-p_2q_2+k_2)}\\
p_1&=\frac{4q_1^2k_2^2+4q_1q_2k_1k_2+q_2^2k_1^2-4q_2^2}{2k_2(2q_1^3k_2+q_1^2k_1q_2-2q_1k_2q_2^2-k_1q_2^3)}.
\end{split}
\end{equation}
Let us notice that the equations \eqref{exactsol} depend on a particular solution $(q_2,p_2)$ and two constants of integration $(k_1,k_2)$ which are related to initial conditions. 

Let us show now the graphs and values of the initial conditions for which the solution reminds us of sigmoid behavior, which is the expected growth of $\rho(t)$. As particular solution for $(q_2,p_2)$, we have made use of particular solution 2 given in Figure \ref{PS2} through its corresponding values of $q,p$ through the change of variables $q=\langle \rho \rangle $ and $p=1/\sigma$.

\begin{figure}[h!]
\centering
\includegraphics[width=0.5\textwidth]{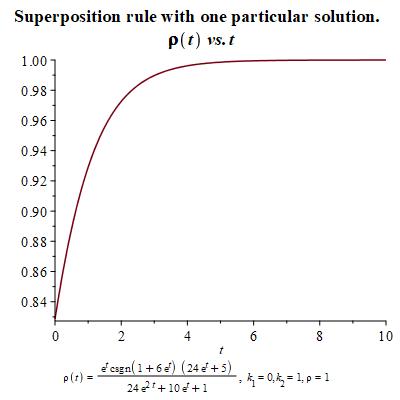}
\label{SR1notlin}
\caption{Superposition rule for exact solution}
\end{figure}

Notice that we have not included the error graph since it gives a constant zero graph because $k_1\rightarrow 0$

\textbf{Step 3 revisited - linear approximation.} 
Since the solution \eqref{exactsol} is quite complicated,  one may look for a solution of a linearized model. We first employ the following change of coordinates
\begin{equation}\label{changelog}
\{u=\ln{|K_1|}, v=\ln{|K_2|}, w=\ln{|K_3|}\}. 
\end{equation}
In terms of these new variables, the system \eqref{logsys} reads
\begin{equation} \label{sis-s--s}
\frac{du}{1-e^{-2u}}=\frac{dv}{e^{-v}+e^{v-w}-1-e^{-2u}}=\frac{dw}{2e^{-v}-(1+e^{-2u})}.
\end{equation}
One can solve the system above by introducing a linear approximation
\begin{equation}
\begin{split}
1-e^{-2u}&\simeq  2u,  \\
e^{-v}+e^{v-w}-1-e^{-2u}&\simeq   2u-w,\\
2e^{-v}-(1+e^{-2u})&\simeq  2u-2v,
\end{split}
\end{equation}
after which \eqref{sis-s--s} reads
\begin{equation}\label{linearsys}
\frac{du}{2u}=\frac{dv}{2u-w}=\frac{dw}{2u-2v}.
\end{equation}
We can solve now $v$ and $w$ in terms of $u$ and obtain
\begin{equation}
v(u)=k_1u^{-\sqrt{2}/2}+k_2u^{\sqrt{2}/2}+u, \qquad
w(u)=\sqrt{2}\left(k_1u^{-\sqrt{2}/2}-k_2u^{\sqrt{2}/2}\right).
\end{equation}
According to the algorithm for deriving superposition rule, we need to isolate the constants of integration $k_1$ and $k_2$. Hence, the two first integrals read now
\begin{equation}
k_1= u^{\frac{\sqrt{2}}{2}}\big(\sqrt{2}v-\sqrt{2}u+w\big)/{2\sqrt{2}}, \qquad 
k_2=u^{-\frac{\sqrt{2}}{2}}
\big(\sqrt{2}v-\sqrt{2}u-w\big)/2\sqrt{2}.
\end{equation}
Now, if we substitute the coordinate changes in \eqref{changelog} and in \eqref{kchange}, we arrive at the following general solution
\begin{equation}\label{ppiosuprule}
q_1=q_2\exp\Big(-\frac{\ln{(q_2p_2)}}{1+k_1+k_2}\Big),\qquad 
p_1=\frac{1}{q_2}\exp\Big(\frac{\sqrt{2}}{2}\frac{(k_1-k_2)\ln{(q_2p_2)}}{1+k_1+k_2}\Big).
\end{equation}
which can be written as 
\begin{equation}\label{ppiosuprule2}
q_1=q_2\Big(q_2p_2\Big)^{\frac{-1}{1+k_1+k_2}},\qquad  p_1=\frac{1}{q_2}\Big(q_2p_2 \Big)^{\frac{\sqrt{2}}{2}\frac{k_1-k_2}{1+k_1+k_2}}. 
\end{equation}
Notice that the solution depends on a particular solution $(q_2,p_2)$ and two constants of integration $(k_1,k_2)$, as in \eqref{exactsol}.

Let us show now the graphs and values of the initial conditions for which the solution reminds us of sigmoid behavior, which is the expected growth of $\rho(t)$. As particular solution for $(q_2,p_2)$, we have made use of particular solution 2 given in Figure \ref{PS2} through its corresponding values of $q,p$ through the change of variables $q=<\rho>$ and $p=1/\sigma$.

\begin{figure}[h!]
\centering
\begin{subfigure}[b]{0.45\linewidth}
\includegraphics[width=\linewidth]{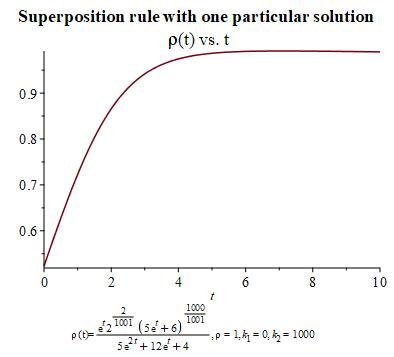}
\end{subfigure}
\begin{subfigure}[b]{0.45\linewidth}
\includegraphics[width=\linewidth]{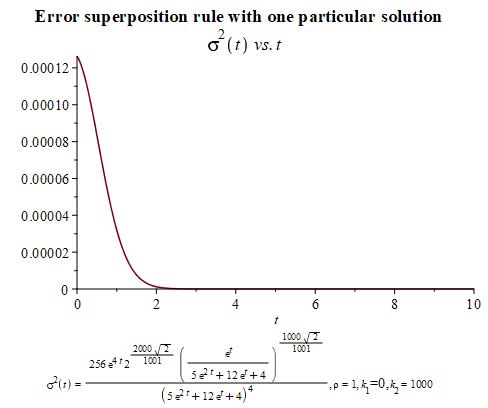}
\end{subfigure}
\caption{Superposition rule for linear approximation}
\label{SR1}
\end{figure}

\section{Lie--Hamilton analysis of the SISf model} \label{Lie-Ham-Sec}

In this section, we shall show that the SISf model \eqref{sismodel3} is a Lie-Hamilton system \cite{Ballesteros, CIMM, LS}. Among the developed methods for Lie--Hamilton systems, we consider a very important recent method for the obtainance of solutions as superposition principles through the Poisson coalgebra method \cite{Ballesteros2,Ballesteros3}. The traditional method for the computation of superposition principles for Lie systems relies in the integration of systems of ordinary or partial differential equations \cite{w1,winter1}, but in the case of Lie--Hamilton systems, the nonlinear superposition rule can be obtained straightforwardly through a Casimir of the Vessiot--Guldberg Lie algebra. We refer to Appendices \ref{App-CB} and \ref{App-PS} for informal summaries of the symplectic and the Poisson geometries, and to Appendix \ref{App-LC} for Lie coalgebras. We refer Appendix \ref{App-PH} for the definition of the symmetric algebra.  

\subsection{Lie--Hamilton systems}~


A Lie-Hamiltonian structure is a triple
$(N,\Lambda,h)$, where $(N,\Lambda)$ stands for a Poisson manifold and $h$
represents a $t$-parametrized family of functions $h_t:N\rightarrow \mathbb{R}$
such that the Lie algebra $\mathcal{H}:={\rm Lie}(\{h_t\}_{t\in\mathbb{R}})$ generated by this family is
finite-dimensional Poisson algebra. 
A time-dependent vector field $X$ is said to possess a Lie-Hamiltonian structure $(N,\Lambda,h)$ if $X_t$ is the Hamiltonian vector field corresponding to $h_t$, for each $t\in\mathbb{R}$,
\begin{equation} \label{identity}
X_t=-\hat{\Lambda} \circ d(h_t ). 
\end{equation}
In this case, $X$ is called a Lie-Hamilton system, and $\mathcal{H}$ is called a Lie--Hamilton algebra for $X$.
Now we examine superposition rules for the case of  Lie--Hamilton systems.

\textbf{Itinerary map for superposition rules of the Lie--Hamilton Systems.}
Let $X$ be a Lie--Hamilton system with a Poisson algebra $\mathcal{H}$ spanned by the linearly independent Hamiltonian functions  $\{ h_1,\dots,h_r \}$. 

\textbf{Step 1.} Consider the injection $\mathfrak{g}\hookrightarrow {\mathcal{H}}$ of a Lie algebra $\mathfrak{g}$ into $\mathcal{H}$ that turns the basis $v_i$ of $\mathfrak{g}$ into the Hamiltonian functions $\phi(v_i):=h_i$ for $i=1,\ldots,r$. Referring to this inclusion, we can define Poisson algebra homomorphisms 
\begin{equation}
D:S(\mathfrak{g}) \to C^\infty(N), \qquad D^{(m)}:  S^{(m)}(\mathfrak{g}) \to C^\infty(N)^{(m)}\subset  C^\infty(N^{m}),
\end{equation}
where $S(\mathfrak{g})$ is the symmetric algebra.

\textbf{Step 2.} 
If $\mathcal{C}$  is a polynomial Casimir  of the Poisson algebra $S(\mathfrak{g})$, say $\mathcal{C}=\mathcal{C}(v_1,\dots,v_r)$, then $D(\mathcal{C})$ is a constant of motion for $X$, and so are the functions defined by
\begin{equation}\label{invA}
F^{(k)}(h_1,\dots,h_r) = D^{(k)}\left[\Delta^{(k)} \left({C( v_1,\dots,v_r)} \right) \right] ,   \qquad k=2,\ldots,m.
\end{equation}
Let us notice, that each $F^{(k)}$ can naturally be considered as a function of $C^\infty(M^m)$ for every $m\geq k$. 

\textbf{Step 3.} From the functions $F^{(k)}$, we can obtain other constants of motion in the form  
\begin{equation}\label{invB}
F_{ij}^{(k)}=S_{ij} ( F^{(k)}   ) , \qquad 1\le  i<j\le  k,\qquad k=2,\ldots,m,
\end{equation}
where $S_{ij}$ is the permutation of variables $x_{(i)}\leftrightarrow
x_{(j)}$ on $M^m$. 

Indeed, since the prolongation $\widetilde X$ is invariant under the permutations $x_{(i)}\leftrightarrow x_{(j)}$, then the $F_{ij}^{(k)}$ are  also $t$-independent constants of motion  for the diagonal prolongations $\widetilde X$ to $M^m$. Let us now illustrate this procedure in the following example.

\begin{example}
Over the plane, consider the following vector fields
\begin{equation}
X_1=\frac{\partial}{\partial x},\quad X_2=\frac{\partial}{\partial y}, \quad X_3=y\frac{\partial}{\partial x}-x\frac{\partial}{\partial y}
\end{equation}
with commutation relations 
$$
[X_1, X_2] = 0,\qquad 	[X_1, X_3]=-X_2, \qquad 	[X_2, X_3] = X_1.
$$

\textbf{Step 1.} With respect to the canonical symplectic structure $\omega={\rm d}x \wedge {\rm d}y$, this corresponds Lie algebra, denoted by $\overline{\mathfrak{iso}}(2)$,  determined by a basis 
\begin{equation}\label{rep}
h_1=y, \qquad h_2=-x, \qquad h_3=\frac 12 (x^2+y^2),\qquad h_0=1
\end{equation}
 satisfying commutation relations   
\begin{equation}
\{h_1,h_2\}_\omega=h_0,\qquad \{h_1,h_3\}_\omega=h_2,\qquad \{h_2,h_3\}_\omega=-h_1,\qquad  \{h_0,\cdot\}_\omega=0 ,
\label{xa}
\end{equation}

\textbf{Step 2.} 
The symmetric Poisson algebra $S\left(\overline{\mathfrak{iso}}(2)\right)$  has a non-trivial  Casimir  invariant given by 
$$
C=v_{3}v_{0}-\tfrac{1}{2}(v_{1}^{2}+v_{2}^{2}).
$$
Choosing the representation given in \eqref{rep}, we obtain a trivial constant of motion on $(x,y)\equiv (x_1,y_1)$
\begin{equation}
\begin{split}
F&=D(C)= \phi(v_3)\phi(v_0)-\tfrac 12 \left( \phi^2(v_1) +\phi^2(v_2) \right) \nonumber\\
& =h_{3}(x_1,y_1)h_{0}(x_1,y_1)-\tfrac{1}{2}\big(h_{1}^{2}(x_1,y_1)+h_{2}^{2}(x_1,y_1)\big)\\
&=\tfrac 12 (x_1^2+y_1^2)\times 1-\tfrac 12 (y_1^2+x_1^2)=0 .
 \end{split}
\end{equation}
Introducing the coalgebra structure in $S\left(\overline{\mathfrak{iso}}(2)\right)$  through the coproduct (\ref{Con}), we obtain nontrivial first integrals.
\begin{equation} \label{intKS}
\begin{split}
  F^{(2)} &= D^{(2) } (\Delta(C) )
\\   
&=  \left(h_3(x_1,y_1)+h_3(x_2,y_2)\right)\left(h_0(x_1,y_1)+h_0(x_2,y_2)\right) \\   
&\qquad\qquad \qquad \qquad-\tfrac 12\Big((\left(h_1(x_1,y_1)+h_1(x_2,y_2)\right)^2+ \left(h_2(x_1,y_1)+h_2(x_2,y_2)\right)^2 \Big)
\\   
&
= \tfrac12
(x_{1}-x_{2})^{2}+\tfrac12(y_{1}-y_{2})^{2}  , 
\\
  F^{(3)}  &=D^{(3) } (\Delta(C) )
\\   
&= \sum_{i=1}^3 h_3(x_i,y_i) \sum_{j=1}^3h_0(x_j,y_j)
-\tfrac 12 \Big(\big( \sum_{i=1}^3 h_1(x_i,y_i)  \big) ^2  + \big( \sum_{i=1}^3 h_2(x_i,y_i)   \big) ^2 \Big)\\
  &=\tfrac12  \sum_{1\le i<j}^3 \big(
(x_{i}-x_{j})^{2}+(y_{i}-y_{j})^{2} \big).
 \end{split}
\end{equation}

\textbf{Step 3.} 
Furthermore, using the property of permutating subindices (\ref{invB}), we find more first integrals
\begin{equation} \label{ff}
\begin{split}
& F_{12}^{(2)}=S_{12} ( F^{(2)}   ) \equiv  F^{(2)}   ,\qquad  F_{13}^{(2)}=S_{13} ( F^{(2)}   )= \tfrac{1}{2} (x_{3}-x_{2})^{2}+\tfrac12(y_{3}-y_{2})^{2} , \\
&  F_{23}^{(2)}=S_{23} ( F^{(2)}   )= \tfrac{1}{2} 
(x_{1}-x_{3})^{2}+\tfrac12 (y_{1}-y_{3})^{2} 
 \end{split}
\end{equation}
Observe that $ F^{(3)}= F_{12}^{(2)}+ F_{13}^{(2)}+ F_{23}^{(2)}$.  We may consider as many first integrals as the number of degrees of freedom of the system in order to integrate it. The combination of these functions leads us to a superposition rule.
 \end{example}

\subsection{The SISf model is Lie-Hamiltonian}~

The SISf epidemic system in \eqref{sismodel3} admits a Hamiltonian formulation as in \eqref{hamiltoniansis}, where the corresponding Hamilton equations are \eqref{sismodel3}.
In this section we will retrieve the Hamiltonian \eqref{hamiltoniansis} using the theory of Lie systems, in particular, the theory of Lie--Hamilton systems. 
To retrieve the Hamiltonian, first we need to prove that equations in \eqref{sismodel3} form a Lie--Hamilton system.  

We have already proven in \eqref{sisliesystem} that \eqref{sismodel3} defines a Lie system. In order to see if it is a Lie--Hamilton system, we first need to check  whether the vector fields in \eqref{sisliesystem} are Hamiltonian vector fields. Consider now the canonical symplectic form $\omega=dq\wedge dp$. It is easy to check that
the vector fields $X_1$ and $X_2$ in \eqref{sisliesystem} are Hamiltonian with respect to the Hamiltonian functions
\begin{equation}\label{hamfunctsis}
h_1=-qp,\quad \quad h_2=-q^2p+\frac{1}{p},
\end{equation}
respectively. It is easy to see that the Poisson bracket of these two functions reads $\{h_1,h_2\}=h_2$. It means that the Hamiltonian functions form a finite dimensional Lie algebra, denoted in the literature as $I^{r=1}_{14A}\simeq \mathbb{R}\ltimes \mathbb{R}$, and it is isomorphic to the one defined by vector fields $X_1, X_2$. The Hamiltonian function for the total system is 
\begin{equation}\label{hamcovid}
h=\rho_0(t)h_1+h_2=-q^2p+\frac{1}{p}-\rho_0(t)qp
\end{equation}
and it is exactly the Hamiltonian function \eqref{hamiltoniansis} proposed in \cite{NakamuraMartinez}.

Lie-Hamilton systems can also be integrated in terms of a superposition rule, as it was explained in the preliminary section. In order to do that, we need to find a Casimir function for the Poisson algebra, but unfortunately, there exists no nontrivial Casimir in this particular case.
 It is interesting to see how a symmetry of the Lie algebra $\{X_1,X_2\}$ commutes with the Lie bracket, i.e. the vector field
\begin{equation}
Z=-\frac{1}{2}\frac{p(C_2p^2q^2+4C_1pq+C_2)}{(pq-1)(pq+1)}\frac{\partial}{\partial p}+\frac{C_1p^4q^4+C_2p^3q^3-C_2pq-C_1}{p(pq-1)^2(pq+1)^2}\frac{\partial}{\partial q}
\end{equation}
fulfills $[X_1,Z]=0,\quad [X_2,Z]=0.$ Notice too that $Z$ is a conformal vector field, that is, 
\begin{equation}
\mathcal{L}_Z\omega=-(C_2/2)\omega.
\end{equation}
Since it is a Hamiltonian system, one would expect that a first integral for $Z$, let us say $f$, would Poisson commute with the Poisson algebra $\{h_1,h_2\}$, since $Z=-\hat{\Lambda}(df)$. Nonetheless, this is not the case unless $f=\text{constant}$. 
This implies that the Casimir is a constant, hence trivial and the coalgebra method can not be directly applied. However, there is a way in which we can circumvent this problem by considering an inclusion of the algebra $I^{r=1}_{14A}$ as a Lie subalgebra of a Lie algebra to another class admitting a Lie--Hamiltonian algebra with a non-trivial Casimir. In this case, we will consider the algebra, denoted by $I_8\simeq \mathfrak{iso}(1,1)$, due to the simple form of its Casimir. If we obtain the superposition rule for $I_8$, we  simultaneously obtain the superposition for $I_{14A}^{r=1}$ as a byproduct.

The Lie--Hamilton algebra $\mathfrak{iso}(1,1)$ has the commutation relations
\begin{equation}\label{basisiso}
\{h_1,h_2\}=h_0,\quad \{h_1,h_3\}=-h_1,\quad  \{h_2,h_3\}=h_2, \quad \{h_0,\cdot\}=0,
\end{equation}
with respect to $\omega=dx\wedge dy$ in the basis 
$\{h_1=y,h_2=-x,h_3=xy,h_0=1\}$. The Casimir associated to this Lie--Hamilton algebra is
$\mathcal{C}=h_1h_2+h_3h_0$. Let us apply
the coalgebra method to this case. Mapping the representation without coproduct, the first
iteration is trivial, i.e., $F=0$. We could usethe second order coproduct and third order coproduct $\Delta^{(2)}$ and $\Delta^{(3)}$, or the second order coproduct $\Delta^{(2)}$ together with the permuting subindices property.
We need three constants of motion, this would be equivalent to integrating 
the diagonal prolongation $\widetilde X$ on $(\mathbb{R}^2)^3$. Using the coalgebra method
and subindex permutation, one obtains
\begin{equation}
\begin{split}
  F^{(2)}&=  
(x_{1}-x_{2})   (y_{1}-y_{2}) =   k_1 , \\
F_{23}^{(2)}&=(x_{1}-x_{3})   (y_{1}-y_{3}) =   k_2 , \\
F_{13}^{(2)}& =(x_{3}-x_{2})   (y_{3}-y_{2}) =   k_3 .
\end{split}
\end{equation}
From them, we can choose two functionally independent constants of motion. Our choice is $F^{(2)}=k_1,F^{(2)}_{23}=k_2$. The introduction of  $k_3$  simplifies the final result, with expression 
\begin{equation}\label{sup2}
\begin{split}
x_1(x_2,y_2,x_3,y_3,k_1,k_2,k_3)&=\frac 12(x_2+x_3) +\frac{k_2-k_1\pm B}{2(y_2-y_3)}   , \\
y_1(x_2,y_2,x_3,y_3,k_1,k_2,k_3)  &=\frac 12(y_2+y_3) +\frac{k_2-k_1\mp B}{2(x_2-x_3)} ,
\end{split}
\end{equation}
where
\begin{equation}
B= \sqrt{ k_1^2+k_2^2+k_3^2-2(k_1k_2+k_1k_3+k_2k_3) }.
\end{equation}

In the case that matters to us, $I_{14A}^{r=1}$, the third constant $k_3$ is a function $k_3=k_3(x_2,y_2,x_3,y_3)$ and $B\geq 0$. Notice though that this superposition rule is expressed in the basis \eqref{basisiso}, therefore, we need the change of coordinates between the present $\mathfrak{iso}(1,1)$ and our problem \eqref{hamfunctsis}.
See that the commutation relation $\{h_1,h_3\}=-h_1$ in \eqref{basisiso} coincides with the commutation relation $\{h_1,h_2\}=h_2$ of our pandemic system \eqref{hamfunctsis}. 
So, by comparison, we see there is a change of coordinates
\begin{equation}\label{changecoord}
x=-qp,\qquad y=q-\frac{1}{qp^2}.
\end{equation}
This way, introducing this change \eqref{changecoord} in \eqref{sup2}, the superposition principle for our Hamiltonian pandemic system reads
\begin{equation}\label{sassaa}
\begin{split}
q&=\frac{\left(\frac{\frac 12 q_2+\frac 12 q_3+(k_2-k_1\pm B)}{(2p_2-2p_3)}\right)^2\left(\frac 12 p_2+\frac 12 p_3+\frac{(k_2-k_1\mp B)}{(2q_2-2q_3)}\right)}{\left(\frac{\frac 12 q_2+\frac 12 q_3+(k_2-k_1\pm B)}{(2p_2-2p_3)}\right)^2-1}
\\
p&=-\frac{\left(\frac{\frac 12 q_2+\frac 12 q_3+(k_2-k_1\pm B)}{(2p_2-2p_3)}\right)^2-1}{\frac 12 q_2+\frac 12 q_3+\frac{(k_2-k_1\pm B)}{(2p_2-2p_3)}\left(\frac 12 p_2+\frac 12 p_3+\frac{(k_2-k_1\mp B)}{(2q_2-2q_3)}\right)}
\end{split}
\end{equation}
Here, $(q_2,p_2)$ and $(q_3,p_3)$ are two particular solutions and $k_1,k_2,k_3$ are constants of integration.

Now, we show the graphics for $<\rho>=q(t)$ and $\sigma^2=1/p^2$ using the two particular solutions in Figure \ref{PS2} and Figure \ref{PS3} provided in the introduction. Notice that we have renamed $c=(k_2-k_1\pm B)$ and $k=(k_2-k_1\mp B)$.

\begin{figure}[h!]
\centering
\begin{subfigure}[b]{0.45\linewidth}
\includegraphics[width=\linewidth]{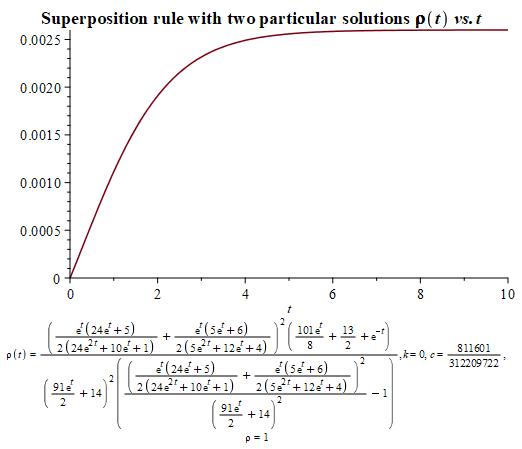}
\end{subfigure}
\begin{subfigure}[b]{0.45\linewidth}
\includegraphics[width=\linewidth]{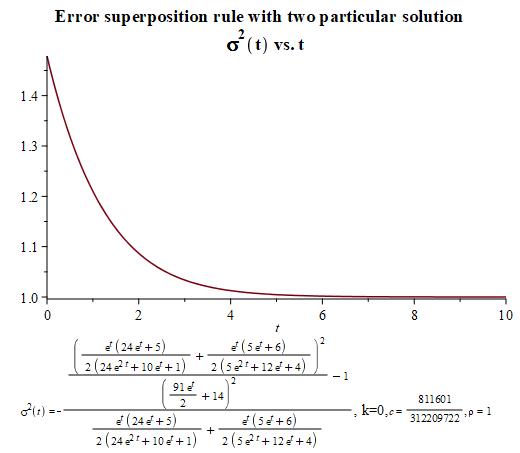}
\end{subfigure}
\caption{Superposition rule with two particular solutions}
\label{SR2}
\end{figure}

\section{A (quantum) deformation of the SISf model} \label{Sec-quant}

A Lie--Hamilton system admits a Poisson--Hopf deformation, see \cite{Ballesteros5}. The interest of Poisson--Hopf deformations of these models resides in the fact that many of the outcome deformed systems happen to be other systems that are already identified in the physics and mathematics literature. This is a highlight, since this basically implies that two different systems scattered in the mathematical physics literature can be merely related by a Poisson--Hopf transformation. This could be a start for the classification of the fuzzy number of integrable systems.

We have interpreted that the SISf model \eqref{sismodel3} is a Lie-Hamilton system. So, in this section we propose a (quantum) deformation of the SISf model  \cite{Ballesteros2,Ballesteros4,BHLS,CLS,LS} which does not rely on the
complicated functional analysis considerations for unbounded quantum operators \cite{Michor}. 
The introduction of the quantum approach will account for the existence of a 
possible interaction of the SISf model with a heat bath, which is effectively an infinite pool of thermal energy at a given constant temperature. This could correspond with centrally heated buildings. We refer Appendix \ref{App-PH} for some details on Poisson-Hopf algebras.

\subsection{The Poisson–-Hopf algebra deformations of Lie--Hamilton systems}~

To obtain a deformation of the Lie--Hamilton realization of the SISf model we make use of deformed Poisson--Hopf algebras. Following \cite{Ballesteros5}, we summarize the procedure for the planar systems as follows: 

\textbf{Itinerary map to arrive at a deformation of a Lie-Hamilton system.} 
Consider a Lie--Hamilton system given by a time-dependent vector field $X_t\in \mathbb{R}^2$ describing a Lie system. There is a corresponding Lie algebra of Hamiltonian functions $\text{Lie}(\{h_t\}_{t\in \mathbb{R}})$ satisfying \eqref{identity} that defines the Lie system as a Lie--Hamilton system.

\textbf{Step 1.} Deform the space $C^{\infty}(\{h_{t_z}\}^*_{t\in \mathbb{R}})$ with a deformation parameter $z\in\mathbb R$ and accordingly the Poisson bracket
\begin{equation}
\{h_{z,i},h_{z,j}\}_{\omega}= F_{z,ij}(h_{z,1},\dots,h_{z,l }).
\label{zab}
\end{equation}
Here, $F_{z,ij}$ are  certain smooth functions also depending smoothly on the deformation parameter $z$ and  such that
\begin{equation}
\lim_{z\to 0} h_{z,i}=h_i ,\qquad \lim_{z\to 0}\nabla h_{z,i}=\nabla h_i,  \qquad   \lim_{z\to 0}   F_{z,ij}(h_{z,1},\dots,h_{z,l }) =\sum_{k=1}^l c_{ij}^k h_k, 
\label{zac}
\end{equation}
where  $\nabla$ is the gradient.

\textbf{Step 2.} Deform the vector fields ${X}_{z,i}$ as
\begin{equation} \label{contract2}
\iota_{{X}_{z,i}}\omega :={\rm d}h_{z,i}.
\end{equation}

\textbf{Step 3.} The deformed total dynamics is encoded in 
\begin{equation}
{X}_z:=\sum_{i=1}^lb_i(t){X}_{z,i}.
\end{equation}

Notice that when the parameter goes to the zero, we have 
\begin{equation}\label{zad}
  \lim_{z\to 0} \{h_{z,i},h_{z,j}\}_{\omega}=\{h_i,h_j\}_\omega ,\qquad \lim_{z\to 0} {X}_{z,i}= {X}_i,
\end{equation}
and the deformed dynamics reduces to the initial one. 

\subsection{The deformed SISf Model}~
 
For the SISf model \eqref{sismodel3}, we start with the Vessiot--Guldberg algebra \eqref{sisliesystem} labelled as $I_{14A}^{r=1}$. To obtain a deformation of a Lie algebra $I_{14A}^{r=1}$, we need to rely on a bigger Lie algebra, in this case, we make use of $\mathfrak{sl}(2)$. To this end, consider the vector fields $X_1$ and $X_2$ in \eqref{sisliesystem-}, and let $X_3$ be a vector field given by \begin{equation}
    X_3:=\frac{p^2q^2(-2p^2q^2+c+6)+c}{2(p^2q^2-1)^2}\frac{\partial}{\partial q}-\frac{p^3q(c+2)}{(p^2q^2-1)^2}\frac{\partial}{\partial p},
\end{equation} where $c\in \mathbb{R}$. Then, $\{X_1,X_2,X_3\}$ span a Vessiot--Lie algebra $V$ isomorphic to $\mathfrak{sl}(2)$ that satisfies the following commutation relations \begin{equation}\label{sl2commrule}
[X_1,X_2]=X_2,\qquad [X_1,X_3]=-X_3,\qquad [X_2,X_3]=2X_1 .
\end{equation}
This vector field $X_3$ admits a Hamiltonian function, say $h_3$,  with respect to the canonical symplectic form on $\mathbb{R}^2$, so that we have the family
 \begin{equation}
    h_1=-qp,\qquad \quad h_2=\frac{1}{p}-q^2p, \qquad h_3=\frac{2p^3q^2+c}{2-2p^2q^2}.
\end{equation} Hence, $\{h_1,h_2,h_3\}$ span a Lie--Hamilton algebra $\mathcal{H}_\omega$; isomorphic to $\mathfrak{sl}(2)$ where the commutation relations with respect to the Poisson bracket induced by the canonical symplectic form $\omega$ on $\mathbb{R}^2$ are given by 
\begin{equation}\label{sl2Poissoncommrule}
\{h_1,h_2\}_\omega=h_2,\qquad \{h_1,h_3\}_\omega=-h_3,\qquad \{h_2,h_3\}_\omega=2h_1.
\end{equation}
Here is the deformation of the model by steps.

\textbf{Step 1.}
Applying the non-standard deformation of $\mathfrak{sl}(2)$ in \cite{Ballesteros5} we  arrive at the Hamiltonian functions 
\begin{equation}\label{gd} 
   h_{z;1}=-shc(2zh_{z;2})qp,  \qquad
 h_{z;2}=\frac{1}{p}-q^2p, \qquad
 h_{z;3}=-\frac{p\left(shc(2zh_{z;2})2q^2p^2+c\right)}{2shc(2zh_{z;2})(q^2p^2-1)} ,
\end{equation}
Here, $shc(x)$ is the cardinal hyperbolic sinus function. Accordingly, the Poisson brackets are computed to be
\begin{equation}\label{gb2}
\begin{gathered}
\{h_{z;1},h_{z;2}\}_\omega=shc(2zh_{z;2})h_{z;2},\qquad 
 \{h_{z;2},h_{z;3}\}_\omega=2h_{z;1},\\[2pt]
 \{h_{z;1},h_{z;3}\}_\omega=-cosh(2zh_{z;2})h_{z;3},
\end{gathered}
\end{equation}

\textbf{Step 2.} The vector fields $X_{z;1}$ and $X_{z;2}$ associated to the Hamiltonian functions $h_{z;1}$ and $h_{z;2}$ exhibited in \eqref{gd} are
\begin{equation}
\begin{split}
X_{z,1}&=\frac{cosh\left(2z(\frac{1}{p}-q^2p)\right)}{(p^2q^2-1)^2}\left[(1-p^4q^4) \frac{\partial}{\partial q}+(2p^5q^4-p^3q^2)\frac{\partial}{\partial p}\right]\\ &\qquad \qquad +\frac{shc\left(2z(\frac{1}{p}-q^2p)\right)}{(p^2q^2-1)}\left[ q \frac{\partial}{\partial q}-p(p^2q^2+1)\frac{\partial}{\partial p}\right],\\ 
X_{z,2}&=\left(-q^2-\frac{1}{p^2}\right)\frac{\partial}{\partial q}+2qp\frac{\partial}{\partial p}.
\end{split}
\end{equation}
We do not write explicitly the expression of the vector field $X_{z;3}$ because it does not play a relevant role in our system. The deformed vector fields keep the commutation relations
\begin{equation}\label{com2}
\left[X_{z;1},X_{z;2}\right]=cosh\left(2z\left(\frac{1}{p}-q^2p\right)\right) \, X_{z;2}.
\end{equation}

\textbf{Step 3.}  The total Hamiltonian function for the deformed model is
\begin{equation}\label{defham}
h_z=\rho(t)h_{z;1}+h_{z;2}=-\rho(t) shc(2zh_{z;2})qp+\frac{1}{p}-q^2p.
\end{equation} 
so that the deformed dynamics is computed to be
\begin{equation}\label{dssis}
\begin{split}
    \frac{dq}{dt}&=\left(\frac{cosh\left(2z(\frac{1}{p}-q^2p)\right)}{(p^2q^2-1)^2}(1-p^4q^4) +\frac{shc\left(2z(\frac{1}{p}-q^2p)\right)}{(p^2q^2-1)}\ q \right)\rho_0(t)-q^2-\frac{1}{p^2},\\
    \frac{dp}{dt}&=\left(\frac{cosh\left(2z(\frac{1}{p}-q^2p)\right)}{(p^2q^2-1)^2}(2p^5q^4-p^3q^2)-p\frac{shc\left(2z(\frac{1}{p}-q^2p)\right)}{(p^2q^2-1)}(p^2q^2+1)\right)\rho_0(t)-2qp.
\end{split}
\end{equation}
This system describes a family of z-parametric differential equations that generalizes the SISf model \eqref{sismodel3}, where the demographic interaction and both rates allow a more realistic representation of the epidemic evolution. According to the kind of deformation, this may be called a quantum family SISf model. Note that the SISf model can be recovered in the limit when $z$ tends to zero.  

\textbf{Constants of motion.} For the present case, the constants of motion defined in \eqref{invA} are computed to be \begin{equation}
    F^{(1)}=\frac{c}{4},\qquad F^{(2)}=\left(h_{2}^{(1)}+h_{2}^{(2)}\right)\left(h_{3}^{(1)}+h_{3}^{(2)}\right)-\left(h_{1}^{(1)}+h_{1}^{(2)}\right)^2,
\end{equation}
after the quantization, the latter one becomes 
\begin{equation}\label{constant1}
   F_z^{(2)}= shc\left(2zh_{z;2}^{(2)}\right)h_{z;2}^{(2)}h_{z;3}^{(2)}-\left(h_{z;1}^{(2)}\right)^2,
\end{equation}  where $h_{z;j}^{(2)}:=D_z^{(2)}(\Delta_z(v_j))$. This coproduct $\Delta_z$ can be described as a follows \begin{equation*}
\Delta_z(v_2)=  v_2 \otimes 1+1\otimes v_2 , \qquad
\Delta_z(v_j)=v_j \otimes e^{2 z v_2} + e^{-2 z v_2} \otimes v_j   ,\qquad  j=1,3.
\end{equation*} 
More explictly, using the expressions given in \eqref{gd}, we have
\begin{equation}
\begin{split}
h_{z;j}^{(2)}=&h_{z;j}(q_1,p_1)e^{2zh_{z;2}(q_2,p_2)}+h_{z;j}(q_2,p_2)e^{-2zh_{z;2}(q_1,p_1)}, \qquad j=1,3 \\ h_{z;2}^{(2)}=&h_{z;2}(q_1,p_1)+h_{z;2}(q_2,p_2).
\end{split}
\end{equation}

So, to retrieve another constant of motion we can apply the trick of permuting indices.
Then, here we have a second constant of motion, writing it implicitly,
\begin{equation}\label{constant2}
  F_{z,(23)}^{(2)}= shc\left(2zh_{z;2 (23)}^{(2)}\right)h_{z;2(23)}^{(2)}h_{z;3(23)}^{(2)}-\left(h_{z;1 (23)}^{(2)}\right)^2,
  \end{equation}
  where the subindex $(23)$ means that the variables $(q_2,p_2)$ are interchanged with $(q_3,p_3)$ when they appear in the deformed Hamiltonian functions $h_{z;j}$ and
  \begin{align*}
h_{z;j (23)}^{(2)}=&h_{z;j}(q_1,p_1)e^{2zh_{z;2}(q_3,p_3)}+h_{z;j}(q_3,p_3)e^{-2zh_{z;2}(q_1,p_1)}, \qquad j=1,3 \\ h_{z;2 (23)}^{(2)}=&h_{z;2}(q_1,p_1)+h_{z;2}(q_3,p_3).
\end{align*}
In \eqref{constant1}, we have
\begin{equation}
\begin{split}
h_{z;2}(q_1,p_1)&=-shc(2zh_{z;2})q_1p_1,\quad h_{z;2}=\frac{1}{p_1}-q_1^2p_1,\quad h_{z;3}=-\frac{p_1\left(shc(2zh_{z;2})2q_1^2p_1^2+c\right)}{2shc(2zh_{z;2})(q_1^2p_1^2-1)} \\
h_{z;2}(q_2,p_2)&=-shc(2zh_{z;2})q_2p_2,\quad h_{z;2}=\frac{1}{p_2}-q_2^2p_2,\quad h_{z;3}=-\frac{p_2\left(shc(2zh_{z;2})2q_2^2p_2^2+c\right)}{2shc(2zh_{z;2})(q_2^2p_2^2-1)} 
\end{split}
\end{equation}
whilst in \eqref{constant2}
\begin{equation}
\begin{split}
h_{z;2}(q_1,p_1)&=-shc(2zh_{z;2})q_1p_1,\quad h_{z;2}=\frac{1}{p_1}-q_1^2p_1,\quad h_{z;3}=-\frac{p_1\left(shc(2zh_{z;2})2q_1^2p_1^2+c\right)}{2shc(2zh_{z;2})(q_1^2p_1^2-1)} \\
h_{z;2}(q_3,p_3)&=-shc(2zh_{z;2})q_3p_3,\quad h_{z;2}=\frac{1}{p_3}-q_3^2p_3,\quad h_{z;3}=-\frac{p_3\left(shc(2zh_{z;2})2q_3^2p_3^2+c\right)}{2shc(2zh_{z;2})(q_3^2p_3^2-1)} 
\end{split}
\end{equation}
If we set these two first integrals equal to a constant, $F_{z,(23)}^{(2)}=k_{23}$ and $F_z^{(2)}=k_{12}$, with $k_{23},k_{12}\in \mathbb{R}$, one is able to retrieve a superposition principle for $q_1=q_1(q_2,q_3,p_2,p_3,k_{12},k_{23})$ and $p_1=p_1(q_2,q_3,p_2,p_3,k_{12},k_{23})$.
Notice that here $(q_2,p_2)$ and $(q_3,p_3)$ are two pairs of particular solutions and
$k_{12},k_{23}$ are two constants over the plane to be related to initial conditions.

\section{Conclusions}

Here we present a summary of our new results, we conclude how they agree with the experimental data and how they clash with other models and results by other authors. Let us list our major results.

\textbf{Our results.}

\begin{itemize}
\item We have achieved a more general solution \eqref{solsismodel3} 
for the SISf system, a SIS model permitting fluctuations, presented in \eqref{sismodel3} than the one provided in \cite{NakamuraMartinez}.

\item We have limited the range of $C_1$ and $C_2$ in the solution 
\eqref{solsismodel3} for system \eqref{sismodel3}, so that we obtain sigmoid-type or hyperbolic-type like behavior, which is the one that we could expect for the growth of infected individuals $\rho(t)$. This range is $3C_2\leq C_1\leq 10^{14}C_2$. For simplification, we choose $C_2=1$. If one exceeds these limits, it is easy to spot singularities and unexpected behavior of the function. This is precisely what we have realized as we plotted the solutions given in \cite{NakamuraMartinez}. We are still unsure whether their solution is an actual solution or if their choice of initial conditions are not the ideal to overcome singularities and find the real regime.

\item The error $\sigma^2$ decays with time, as the density of infected individuals stabilizes as constant. This is what one expects intuitively.

\item We have been able to retrieve general solutions for equations \eqref{sismodel3} making use of geometric methods. The theory of Lie systems has permitted us to find a general solution in terms of a particular solution and a challenging choice of the constants of integration $(k_1,k_2)$, in both cases, the linear approximation and the unlinearized version.

\item We have been able to retrieve a general solution for \eqref{sismodel3} making use of the theory of Lie--Hamilton systems and it is expressed in terms of two particular solutions and a very challenging choice of two constants $(c,k)$. One can check the choices of constants for every case in the legends of the graphics.

\item It is interesting to notice that using the geometric methods of the Lie--Hamilton theory, one retrieves the introduced hamiltonian \eqref{hamiltoniansis} by \cite{NakamuraMartinez} as performed in \eqref{hamcovid}.

\item Furthermore, we have generalized this hamiltonian \eqref{hamiltoniansis} by introducing a deformation parameter, know as the quantum deformation parameter using the Poisson--Hopf algebra method. The deformed hamiltonian \eqref{defham} gives rise to a new integrable SIS model in which one includes a quantum deformation parameter \eqref{dssis}.
\end{itemize}

\textbf{Relation with Covid19 pandemic.} One may wonder how the current pandemic of COVID19 could be related to a SISf-pandemic model. The SISf model is a very first approximation for a trivial infection process, in which there is only two possible states of the individuals in the population: they are either infected or susceptible to the infection. Hence, this model does not provide the possibility of adquiring immunity at any point. It seems that COVID19 provides some certain type of immunity, but only to a thirty percent of the infected individuals, hence, a SIR model that considers ``R" for recuperated individuals (not susceptible anymore, i.e., immune) is not a proper model for the current situation. One should have a model contemplating immune and nonimmunized individuals. Unfortunately, we are still in search of a stochastic Hamiltonian model including potential immunity and nonimmunity. 

\textbf{Future work.}
As future work, we would like to extend our study to more complicated comparmental models, although at a first glance we have not been able to identify more Lie systems, at least in their current PDE form. We suspect that the Hamiltonian description of these compartmental models could nonetheless behave as a Lie system, as it has happened in our presented case. This shall be part of our future endeavors. Moreover, one could inspect in more meticulous detail how the solutions of the quantumly-deformed system \eqref{dssis} recover the nondeformed solutions when the introduced parameter tends to zero. We need to further study how this precisely models a heat bath and if this new integrable system could correspond to other models apart from infectious models. We would like to figure out whether it is possible to modelize subatomic dynamics with the resulting deformed hamiltonian \eqref{defham}.
 
Apart from that, there exists a stochastic theory of Lie systems developed in \cite{Ortega} that could be another starting point to deal with compartmental systems. In the present work we were lucky to find a theory with fluctuations that happened to match a stochastic expansion, but this is rather more of an exception than a rule. Indeed, it seems that the most feasible way to propose stochastic models is using the stochastic Lie theory instead of expecting a glimpse of luck with fluctuations.

As we have stated, finding particular solutions is by no means trivial. The analytic search is a very excruciating task. We think that in order to fit particular solutions in the superposition principle, one may need to compute these particular solutions numerically. Some specific numerical methods for particular solutions of Lie systems can be devised in \cite{piet}.

\section{Acknowledgment}

OE is greateful Prof. Serkan S\"utl\"u for his kind interest and visionary comments especially on Hopf algebras. CS acknowledges her new position at Universidad Polit\'ecnica de Madrid. EFS acknowledges a
fellowship (grant CT45/15-CT46/15) supported by the Universidad Complutense de
Madrid and was partially supported by grant MTM2016-79422-P (AEI/FEDER, EU)

\pagebreak
\appendix 

\section{}  \label{fund}

\subsection{Lie Algebras} \label{App-LA}~

Let $A$ be a vector space over a field (in this work $\mathbb{R}$). Consider the multiplication map $m$ and a unit element map $\iota$, given by
\begin{equation} \label{algebra}
m:A\otimes A \longrightarrow A, \qquad \iota: \mathbb{R} \longrightarrow A. 
\end{equation}
We say that the triple $(A,m,\iota)$ is an algebra. 
This algebra is said to be associative if the multiplication is associative, i.e., 
\begin{equation}
m(x,m(x',x''))= m(m(x,x'),x'')
\end{equation}  
for all $x$, $y$, and $z$ in $A$. Otherwise it is referred to as a  nonassociative algebra. A mapping between two algebras is called homomorphism if it respects the multiplicatios and unit from one algebra to the other. 

An algebra is called a Lie algebra, denoted by $\mathfrak{g}$, if the multiplication is skew-symmetric and satisfies the Jacobi identity \cite{AzIz98,Gil}. In this case, the multiplication operation is a Lie bracket $[\bullet,\bullet]$. The Jacobi identity is  
\begin{equation} \label{Jac-id}
[[x,x'],x'']+[[x',x''],x]+[[x'',x],x']=\textbf{0}
\end{equation}
for all $x$, $x'$ and $x''$ in $\mathfrak{g}$. If this identity is fulfilled, then the nonassociativity of the multiplication is followed. A mapping between two algebras is called a Lie algebra homomorphism if it preserves the Lie algebra structure.  

Let $B$ be an arbitrary subspace of a Lie algebra $\mathfrak{g}$. We define a new set $[B,B]$ by collecting all possible pairings of the elements in $B$. If the set $[B,B]$ precisely equals $B$, that is, if $B$ is closed under the Lie bracket, we have a Lie subalgebra. In general, we compute the (possible infinite) hierarchy $[B,B],
[B,[B,B]],[[B,B],[B,B]]\dots$ 
consisting of all possible pairings. By continuing in this way, we arrive at a collection that turns out to be a Lie subalgebra of $\mathfrak{g}$. Evidently, it is the smallest Lie subalgebra containing $B$ which we denote ${\rm Lie}(B)$.

\subsection{Tangent Bundle} \label{App-TB} ~

Let $N$ be an $n$-dimensional manifold (a locally Euclidean space).  
A differentiable curve on $N$ through the point $x$ is a function $\gamma :
\mathbb{R} \rightarrow Q$ with, say, $\gamma \left( 0\right) =x$. Let us define now an equivalence relation in the set of differentiable curves passing through $x$. Two curves $\gamma $ and $%
\tilde{\gamma}$ are equivalent if they take the same value at $x$ and if the
directional derivative of functions along them at $x$ are the same, namely,
\begin{equation}
\gamma \left( 0\right) =\tilde{\gamma}\left( 0\right) , \qquad \left.\frac{d}{dt}\right|_{t=0}(f\circ\gamma )\left( t\right) =\left.\frac{d}{dt}\right|_{t=0}\left( f\circ \tilde{\gamma}\right) \left(
t\right)
\end{equation}
for all functions $f:N\rightarrow 
\mathbb{R}$. A tangent vector $v(x)$ at $x$ is an equivalence class
of curves at $x$. The set of all equivalence classes of
vectors, that is, the set of all tangent vectors at $x$ is  tangent space $T_{x}N$ at $x$ in $N$. It is possible to show that $T_{x}N$ admits an $n$-dimensional vector space structure. 

The union of all tangent spaces $T_{x}N$ in $x\in N$ is 
\begin{equation}
TN=\bigsqcup\limits_{x\in N}T_{x}N
\end{equation}
and it is a $2n$-dimensional manifold that is called tangent
bundle of $N$. From the point of view of applications, if a manifold is the configuration space of a physical system, then, the tangent bundle is the velocity phase space of the system. That is, it consists of all possible positions and all possible velocities.   
There is a projection, called  tangent fibration , $\tau_N$ mapping a tangent vector to its base point
\begin{equation}
\tau_N:TN\longrightarrow N, \qquad v(x)\mapsto x.
\end{equation}

\textbf{Vector fields.} A section of the tangent fibration is a  vector field  $X$ on $N$ 
mapping a vector to each point in $N$, that is a map $Y:N\rightarrow TN$
such that $\tau _{N}\circ Y:id_{N}$ where $id_{N}$ is the identity map on $N$. The set of vector fields $\mathfrak{X}(N)$ on a manifold $Q$ admits a module structure over the ring of functions. 
\begin{equation}
\mathfrak{X}(N)=\{Y:N\mapsto TN \quad \vert\quad  Y\circ \tau_N =id_N \}.
\end{equation}
An integral curve of a vector field $X$ with initial condition $x$, is a curve $\gamma$ passing through $\gamma(0)=x$ so that 
\begin{equation} \label{int-curve}
\left.\frac{d}{dt}\right|_{t=0}\gamma(t)=Y(x).
\end{equation}
From the theorem of existence and uniqueness of solutions for ODEs,  we know that for every $x$ there exists a unique integral curve $\gamma$ satisfying \eqref{int-curve}. The flow of $Y$ is a smooth one parameter group of diffeomorphisms defined by means of integral curves as follows 
 \begin{equation}
 \gamma_{t}:N \longrightarrow N, \qquad x \mapsto \gamma(t),
 \end{equation}
 where $\gamma(0)=x$. 
This realization of vector fields enables us to define the {directional derivative} of a function $f$ along a vector field $Y$ as 
\begin{equation}
Y:\mathcal{F}(N)\longrightarrow \mathcal{F}(N),\qquad f\mapsto Y(f)=\left.\frac{d}{dt}\right|_{t=0} (f \circ \gamma)(t)
\end{equation}
where $\gamma$ is an integral curve of $Y$. Notice that, if $f$ is a constant of motion, then, $Y(f)=0$. 

\textbf{Lie Algebras.} The space of all vector fields, denoted by $\mathfrak{X}(N)$, is a Lie algebra if it is endowed with a Lie bracket. For any two vector fields $Y_1$ and $Y_2$ on the base manifold $N$, the Lie bracket reads 
\begin{equation}
[Y_1,Y_2](f)=Y_1(Y_2(f))-Y_2(Y_1(f)),
\end{equation}
for all smooth functions $f$ on $N$. Here, the notation $Y(f)$ is simply the directional derivative of the function $f$ in the direction of $Y$. 
A distribution generated by a finite set of vector fields, say $Y_1$, ... $Y_r$, is the collection of subspaces $\mathcal{D}_x$, generated by the vectors $Y_1(x)$, ... $Y_r(x)$, in all points. It is called an involutive distribution if the Lie algebra is closed for the set $Y_1$, ... $Y_r$. 

\textbf{Local realization in $2D$.} Assume now that $N$ is a two dimensional manifold with local coordinates $(q,p)$, then the tangent bundle $TN$ admits the induced coordinates $(q,p;\dot{q},\dot{p})$. In this case, a vector field is written as 
\begin{equation} \label{vf-loc}
X=k(q,p)\frac{\partial}{\partial q}+g(q,p)\frac{\partial}{\partial q}
\end{equation}
where $k$ and $g$ are real valued functions on $N$. In this coordinate system, the dynamical equations governed by $X$ determine a system of nonautonomous ordinary differential equations
\begin{equation} \label{2D-auto}
\dot{q}=k(q,p), \qquad \dot{p}=g(q,p).
\end{equation}
An integral curve to the vector field $X$ in \eqref{vf-loc} is a solution of the system \eqref{2D-auto} given by $q=q(t)$ and $p=p(t)$. 

\subsection{Cotangent Bundle - Canonical Symplectic Space} \label{App-CB} ~

A tangent space $T_qQ$ at point $q$ admits a (finite dimensional) vector space structure. So, there exists the linear algebraic dual of $T_qQ$, which we denote by $T_q^*Q$.  We call $T_x^*N$ a cotangent space. By collecting all of them, we define the cotangent bundle
\begin{equation}
T^*Q=\bigsqcup_{q\in Q} T_q^*Q.
\end{equation}
There is a projection $\pi _{Q}$ from $T^{\ast }Q$ to $Q$, mapping a covector $\alpha(q)$ to its base point $q$. Sections $\theta $ of the cotangent bundle are one-forms on $Q$
satisfying $\pi _{Q}\circ \theta =id_{Q}$.

\textbf{$2D$ Symplectic Manifold.} Assume now that $N$ is a two dimensional planar manifold with  coordinates $(q,p)$. For this case, we can understand $N=T^*Q$ for some one-dimensional $Q$. There is a canonical symplectic two-form $\omega$ on $N$. A symplectic two-form is a closed, skew-symmetric and nondegenerate bilinear mapping taking two vector fields to the space of scalars, i.e., $\omega$ takes two vector fields $Y_1$ and $Y_2$ on $N$ to a unique function. In this case, the symplectic two-form is computed to be
\begin{equation}
\omega=dq\wedge dp.
\end{equation} 
It is the nondegeneracy of the symplectic two-form what makes the Hamilton equations \eqref{Ham-Eq} be realized in their local form \eqref{loc-Ham-Eq}. It assigns uniquely a vector field $X_h$ to a chosen (Hamiltonian) function $h$. From the point of view of applications in classical mechanics, if the Hamiltonian function $h$ is the total energy, then $X_h$ describes the motion.

\subsection{Poisson Algebras} \label{App-PS}~

An associative algebra $(A,m,\iota)$ is called Poisson if it admits a (Poisson) bracket $\{\bullet,\bullet\}$ satisfying the Jacobi identity, so that the triple $(A,\{\bullet,\bullet\},\iota)$ becomes a Lie algebra, and the Leibniz identity is satisfied \cite{We98}.
\begin{equation} \label{Leib-id}
\{ m(x,y),z\}=m(x,\{y,z\})+m(\{x,y\},z),
\end{equation}
  
We denote a Poisson algebra by a quadruple $(A,m,\iota,\{\bullet,\bullet\})$. An element of the algebra is called Casimir, and denoted by $C$, if it commutes with all other elements under the Poisson bracket, that is if 
\begin{equation}
\{C,x\}=0, \qquad \forall x \in A.
\end{equation}

\textbf{Poisson manifolds.} 
A manifold, say $P$, is called Poisson if it is equipped with a skew-symmetric (Poisson) bracket $\{\bullet,\bullet\}$ on the space of smooth functions on $P$ satisfying both the Jacobi identity  \eqref{Jac-id} and the Leibniz identity \eqref{Leib-id}, see \cite{We83}.   
A Hamiltonian vector field $X_h$ on $N$ associated with a smooth (Hamiltonian) function $h$, is a unique vector field such that the following identity holds
\begin{equation} \label{X_h}
X_h(g):=\{g,h\}.
\end{equation}
The Jacobi identity for the Poisson bracket therefore entails that $h\mapsto X_h$ is a Lie algebra anti-homomorphism between a
Poisson algebra $C^{\infty}(N)$ endowed with a Poisson bracket and space of vector fields $\mathfrak{X}(N)$ endowed with a Lie bracket. Notice that every symplectic manifold is a Poisson manifold  and in this case the Hamilton equation \eqref{Ham-Eq} takes the form of \eqref{X_h}. 

\textbf{Lie-Poisson bracket.} Let $\mathfrak{g}$ be a Lie algebra, and $\mathfrak{g}^*$ be its linear algebraic dual space. $\mathfrak{g}^*$ is  a Poisson manifold, known as a Lie-Poisson manifold with the Lie-Poisson bracket 
\begin{equation} \label{LP-bracket}
\{h,g\}(z)=-\big \langle z,\big[\frac{\delta h}{\delta z},\frac{\delta g}{\delta z} \big] \big \rangle,
\end{equation}
for any $z$ in $\mathfrak{g}^*$, and any two function(als) $h$ and $g$ on $\mathfrak{g}^*$, \cite{MaRa}. Here, the bracket on the right hand side is the Lie algebra bracket on $\mathfrak{g}$. Let us note that ${\delta h}/{\delta z}$ and ${\delta g}/{\delta z}$ denote the partial (for infinite dimensional cases Fr\'{e}chet) derivatives of the function(al)s. Under the assumption of the reflexivity, they are elements of $\mathfrak{g}$.

\textbf{Poisson bivectors.} Being a derivation for each factor, a Poisson structure determines a unique bivector field $\Lambda\in 
\Gamma(\wedge^2 TN)$, that we call a Poisson bivector, such that
\begin{equation}\label{ForPoi}
\Lambda(dh,dg):=\{h,g\}.
\end{equation}
As a manifestation of the Jacobi identity, the Schouten--Nijenhuis bracket of the bivector $[\Lambda,\Lambda]_S$ equals zero.  Conversely, every bivector field
 $\Lambda$ on $N$ satisfying the null condition
gives rise to a Poisson structure. A bivector induces a unique bundle morphism from the space of one-forms $\Omega^1(N)$ to the space of vector fields $\mathfrak{X}(N)$ on $N$, that is, 
\begin{equation}
\hat\Lambda:\Omega^1(N)\rightarrow \mathfrak{X}(N), \qquad \langle \beta, \hat\Lambda(\alpha) \rangle:=\Lambda(\alpha, \beta).
\end{equation}
Referring to this mapping, we define a Hamiltonian vector field \eqref{X_h}, as follows 
\begin{equation}
X_h=-\hat \Lambda \circ dh,
\end{equation}
where $dh$ is the (exterior) derivative of $h$.  
 The space of Hamiltonian vector fields induces an integrable generalized distribution $\mathcal{F}^\Lambda$ on $N$ associated to $\Lambda$. Pointwisely, each fiber $\mathcal{F}^\Lambda_x$ of this distribution is defined to be the image space of the mapping $\hat\Lambda$, that is 
  $\{X_x\mid X \in {\rm Im}\, \hat \Lambda\}$. Here, the leaves are symplectic manifolds with respect to the restrictions of $\Lambda$ \cite{We83}.


\subsection{Lie coalgebras} \label{App-LC}~

In order to arrive the formal definition of a coalgebra, one simply reverse the directions of the (multiplication and unit) arrows in \eqref{algebra}. Accordingly, a vector space $A$ is called a coalgebra if it admits a comultiplication and a counit given by
\begin{equation} \label{coalgebra}
\begin{split}
\Delta&:A \longrightarrow A\otimes A, \qquad  
\Delta z = z\nsb{1} \otimes z\nsb{2}, 
\\
\epsilon &: A\longrightarrow
\Bbb R,\qquad   \epsilon(z)=1
\end{split}
\end{equation} 
respectively. We ask that these operations must satisfy 
the relation
\begin{equation}
({\rm Id}\otimes\epsilon)\Delta (z)=(\epsilon\otimes {\rm Id})\Delta (z)= z, \qquad \forall z \in A,
\end{equation}
where ${\rm Id}$ is being the identity mapping. We denote a coalgebra by a triple $(A,\Delta,\epsilon)$. A mapping is called a coalgebra homomorphism if it respects the coalgebra structures.

\textbf{Lie coalgebras.}
A coalgebra  $A$ is called a Lie coalgebra \cite{Majid-book,Mich80} if the comultiplication, this time called Lie cobracket, 
satisfies the following two conditions
\begin{equation} \label{cond-coalg}
\begin{split}
&z\nsb{1} \otimes z\nsb{2} = - z\nsb{2} \otimes z\nsb{1},\\
&z\nsb{1} %
\otimes z\nsb{2}\nsb{1} \otimes z\nsb{2}\nsb{2} + z\nsb{2}\nsb{1} \otimes z
\nsb{2}\nsb{2} \otimes z\nsb{1} + z\nsb{2}\nsb{2} \otimes z\nsb{1} \otimes z
\nsb{2}\nsb{1} = 0,
\end{split}
\end{equation}
where we code the elements in $A$ from left to the right with numbers, for example, 
\begin{equation}
z\nsb{1} \otimes z\nsb{2}\nsb{1} \otimes z\nsb{2}\nsb{2}: =z\nsb{1} \otimes \Delta z\nsb{2}.
\end{equation}
Notice that the first condition in \eqref{cond-coalg} is \textit{dual} of the skew symmetry whereas the second condition in \eqref{cond-coalg} is \textit{dual} of the Jacobi identity. 

\textbf{Dual of a Lie (co)algebra.} 
The dual of a Lie coalgebra admits a Lie algebra structure. The Lie algebra bracket on the dual space is defined by means of the following equality
\begin{equation}
[\bullet,\bullet]_{A^*}:A^* \otimes A^* \longrightarrow A^*, \qquad \langle [x,x']_{\mathfrak{G}^*}, z \rangle := \Delta z (x,x'),
\end{equation}
where the pairing on the left hand side is the one between $A^*$ and $A$ whereas the pairing on the right hand side is between the tensor products $A^*\otimes A^*$ and $A\otimes A$. 

Inversely, the dual $\mathfrak{g}
^{\ast}$  of a Lie algebra $\mathfrak{g}$ is an immediate example of Lie coalgebras. In this case the cobracket is defined to be
\begin{equation} \label{cobra-Lie-alg}
\Delta:\mathfrak{g}
^{\ast} \longrightarrow \mathfrak{g}
^{\ast} \otimes \mathfrak{g}
^{\ast}, \qquad 
\Delta z\,(x,x'):= \langle z,[x,x']\rangle,
\end{equation}
where the bracket $[\bullet,\bullet]$ on the right hand side is the Lie algebra bracket on $\mathfrak{g}$.

\subsection{Poisson--Hopf algebras} \label{App-PH} ~

Consider a vector space admitting both  an algebra structure $(A,m,\iota)$ and a coalgebra structure $(A,\Delta,\epsilon)$. If these two structures are compatible, that is, either $\Delta$ and $\epsilon$ are algebra homomorphisms or $m$ and $\iota$ are coalgebra homomorphisms, then $A$ is a called a bialgebra \cite{Abe,Chari,Majid-book}. We denote a bialgebra by a quintuple $(A,m,\iota,\Delta,\epsilon)$. 

\textbf{Poisson bialgebra.} Consider a bialgebra $(A,m,\iota,\Delta,\epsilon)$ where there is an associated Poisson bracket  $\{\bullet,\bullet\}$ with $A$. If $\Delta:A\mapsto A\otimes A$ preserves the Poisson structure as well, then, the tuple $(A,m,\iota,\{\bullet,\bullet\},\Delta,\epsilon)$ is a Poisson bialgebra. In this picture the Poisson bracket on the tensor space $A\otimes A$ is defined to be
\begin{equation}
\{(x\otimes x'),(x''\otimes x^\dagger)\}=\{x,x''\}\otimes m(x',x^\dagger) + m(x,x'') \otimes \{x',x^\dagger\}.
\end{equation}

\textbf{Poisson--Hopf algebra.}  A bialgebra $(A,m,\iota,\Delta,\epsilon)$ is called a 
Hopf algebra 
if there exist an  antihomomorphism, known as the {antipode} $\gamma :
A\longrightarrow A$,  such that for every $ a\ \! \in A$ one gets: 
\begin{equation}
m((   {\rm Id}  \otimes \gamma)\Delta (a))=m((\gamma \otimes {\rm Id})\Delta (a))=
\epsilon (a) \iota,
\end{equation}
see \cite{Dr88,MiMo65,Sw69}. We also cite \cite{RaSu12,RaSu19,Su13} for some recent topological and cohomological discussions. 
We denote a Hopf algebra by the tuple $(A,m,\iota,\Delta,\epsilon,\gamma)$. If a bialgebra $(A,m,\iota,\Delta,\epsilon)$ is both a Poisson and a Hopf  algebra, then it is a Poisson--Hopf algebra and it is denoted by the tuple $(A,m,\iota,\{\bullet,\bullet\},\Delta,\epsilon,\gamma)$. 

The space of smooth functions $C^\infty\left(\mathfrak{g}^*\right)$ on the dual of a Lie algebra $\mathfrak{g}$ is a Hopf algebra relative to its natural associative algebra with unit provided that
\begin{equation}
\begin{split}
m(h\otimes g)(z):&=h(z)g(z), \qquad \iota(1)(z):=1, \\
\Delta (f)(z,z'):&=f(z+z')  \epsilon (f):=f(0),\qquad \gamma(f)(z):=f(-z),
 \end{split}
\end{equation}
  for every $z,z'$ in $\mathfrak{g}^*$, and $f,g,h$ in $C^\infty(\mathfrak{g}^*)$. Further, considering the Lie-Poisson bracket \eqref{LP-bracket}, $C^\infty\left(\mathfrak{g}^*\right)$ turns out to be a Poisson--Hopf algebra.

\textbf{Symmetric (co)algebra.} The symmetric algebra $S(\mathfrak{g})$ of a (finite dimensional) Lie algebra 
$\mathfrak{g}$ is the smallest commutative algebra containing $\mathfrak{g}$. To reach such algebra, we do the following. 
The second tensor power $\mathfrak{g}\otimes \mathfrak{g}$ of the Lie algebra is the space of real valued bilinear maps on the dual space. Iteratively, the kth tensor power $\mathfrak{g}^{\otimes k}$ is the space of real valued k-linear maps. Taking the direct sum of the tensor powers of all orders, we arrive at the tensor algebra $\mathfrak{Tg}$ of $\mathfrak{g}$. Here, the multiplication is 
\begin{equation}
\mathfrak{Tg}\times \mathfrak{Tg} \longrightarrow \mathfrak{Tg}, \qquad (v,u)\mapsto v\otimes u.
\end{equation}
We consider a basis $\{x_1,\dots, x_r\}$ of the Lie algebra 
$\mathfrak{g}$.
The space generated by the elements 
\begin{equation}
x_i\otimes x_j-x_j \otimes x_i
\end{equation}
 is an ideal, denoted by $\mathcal{R}$, of the tensor algebra $\mathfrak{Tg}$. The quotient space $\mathfrak{Tg}/\mathcal{R}$ is called a symmetric algebra and denoted by $S(\mathfrak{g})$. 
The elements of $S(\mathfrak{g})$ can be regarded as polynomial functions on $\mathfrak{g}^*$, so, we can endow it with the Lie-Poisson bracket \eqref{LP-bracket} that makes $S(\mathfrak{g})$ a Poisson algebra. One can show that $S(\mathfrak{g})$ can always be endowed with a  coalgebra structure by introducing the comultiplication 
\begin{equation}\label{Con}
{\Delta} : S(\mathfrak{g})\rightarrow
S(\mathfrak{g}) \otimes S(\mathfrak{g}), \qquad 
{\Delta}(x)=x\otimes 1+1\otimes x,\quad \forall  x\in\mathfrak {g}\subset S(\mathfrak{g}),
\end{equation}
which is a Poisson algebra homomorphism. This makes $S(\mathfrak{g})$ a Poisson-Hopf algebra. Furthermore, in the light of the coassociatity condition 
\begin{equation}
\Delta^{(3)}:=(\Delta \otimes {\rm Id}) \circ \Delta=({\rm Id} \otimes \Delta) \circ \Delta,
\end{equation}
we can define the third-order coproduct  \begin{equation}\label{3co}
\Delta^{(3)}: S(\mathfrak{g})\rightarrow
S(\mathfrak{g}) \otimes S(\mathfrak{g})\otimes S(\mathfrak{g}), \qquad {\Delta}^{(3)}(x)=x\otimes 1\otimes 1 +1\otimes x\otimes 1+1\otimes 1\otimes x
\end{equation}
for all  $x\in\mathfrak {g}$, where $\mathfrak g$ is understood as a subset of $S(\mathfrak{g})$. The {$m$th-order coproduct} map 
can be defined, recursively, as  
\begin{equation}\label{copr}
\Delta ^{(m)}:  S(\mathfrak{g})\rightarrow   S^{(m)}(\mathfrak{g}), \qquad {\Delta}^{(m)}:= ({\stackrel{(m-2)-{\rm times}}{\overbrace{{\rm
Id}\otimes\ldots\otimes{\rm Id}}}}\otimes {\Delta^{(2)}})\circ \Delta^{(m-1)},\qquad m\ge 3,
\end{equation}
which, clearly, is also  a Poisson algebra homomorphism.

\textbf{Universal enveloping algebras.} We consider once more the tensor algebra $\mathfrak{Tg}$ of a Lie algebra $\mathfrak{g}$ with 
a basis $\{x_1,\dots, x_r\}$. We now define the space generated by the elements 
\begin{equation}
x_i\otimes x_j-x_j \otimes x_i - [x_i,x_j].
\end{equation}
This space is an ideal, denoted by $\mathcal{L}$, of the tensor algebra. The quotient space $\mathfrak{Tg}/\mathcal{L}$ is called universal enveloping algebra of $\mathfrak{g}$, and it is denoted by $U(\mathfrak{g})$. $U(\mathfrak{g})$ is the biggest associative algebra containing all possible representations of $\mathfrak{g}$. See that, if the Lie algebra bracket on $\mathfrak{g}$ is trivial, then, the universal enveloping algebra $U(\mathfrak{g})$ is equal to the symmetric algebra $S(\mathfrak{g})$.

\end{document}